\documentclass[
superscriptaddress, twocolumn,
 amsmath,amssymb,
 aps,
pra,
]{revtex4-2}

\usepackage{graphicx}% Include figure files
\usepackage{dcolumn}% Align table columns on decimal point
\usepackage{bm}% bold math
\usepackage{hyperref}
\hypersetup{colorlinks,allcolors=blue}
\usepackage{physics}
\usepackage{upgreek}
\usepackage{cleveref}
\usepackage[protrusion=true, expansion=true]{microtype} %better PDF typesetting
\newcommand{\ii}[0]{\textrm{i}}
\newcommand{\ee}[0]{\textrm{e}}
\begin{document}

%\preprint{APS/123-QED}

\title{Multimode physics of the unimon circuit}%

\author{Sasu Tuohino}
    \affiliation{QCD Labs, QTF Centre of Excellence, Department of Applied Physics, Aalto University, P.O. Box 15100, FI-00076 Aalto, Finland}
\affiliation{Nano and Molecular Systems Research Unit, University of Oulu,\\ P.O.~Box 3000, FI-90014 Oulu, Finland}
    
% \altaffiliation{Physics Department, XYZ University.}%Lines break automatically or can be forced with \\
\author{Vasilii Vadimov}
    \affiliation{QCD Labs, QTF Centre of Excellence, Department of Applied Physics, Aalto University, P.O. Box 15100, FI-00076 Aalto, Finland}


\author{Wallace Teixeira}
    \affiliation{QCD Labs, QTF Centre of Excellence, Department of Applied Physics, Aalto University, P.O. Box 15100, FI-00076 Aalto, Finland}

\author{Tommi Malmelin}
    \affiliation{QCD Labs, QTF Centre of Excellence, Department of Applied Physics, Aalto University, P.O. Box 15100, FI-00076 Aalto, Finland}

\author{Matti Silveri}
\affiliation{Nano and Molecular Systems Research Unit, University of Oulu,\\ P.O.~Box 3000, FI-90014 Oulu, Finland}
%\affiliation{Jointly supervised the work.}

\author{Mikko M\"ott\"onen}
    \affiliation{QCD Labs, QTF Centre of Excellence, Department of Applied Physics, Aalto University, P.O. Box 15100, FI-00076 Aalto, Finland}
    \affiliation{VTT Technical Research Centre of Finland Ltd., QTF Center of Excellence, P.O. Box 1000, FI-02044 VTT, Finland}
%\affiliation{Jointly supervised the work.}
    

\begin{abstract}
%{\bf Mikko's first version:} 
We consider a superconducting half-wavelength resonator that is grounded at its both ends and contains a single Josephson junction. Previously this circuit was considered as a unimon qubit in the single-mode approximation where dc-phase-biasing the junction to $\pi$ leads to increased anharmonicity and 99.9\% experimentally observed single-qubit gate fidelity. Inspired by the promising first experimental results, we develop here a theoretical and numerical model for the detailed understanding of the multimode physics of the unimon circuit. To this end, first, we consider the high-frequency modes of the unimon circuit and find that even though these modes are at their ground state, they imply a significant renormalization to the Josephson energy. %We introduce an efficient method how the relevant modes can be fully taken into account
We introduce an efficient method to fully account for the relevant modes and show that unexcited high-lying modes lead to corrections in the qubit energy and anharmonicity. Interestingly, provided that the junction is offset from the middle of the circuit, we find strong cross-Kerr coupling strengths between a few low-lying modes. This observation paves the way for the utilization of the multimode structure, for example, as several qubits embedded into a single unimon circuit.
\end{abstract}

\maketitle

%\tableofcontents

\begin{figure*}[t]
\includegraphics[scale=0.41]{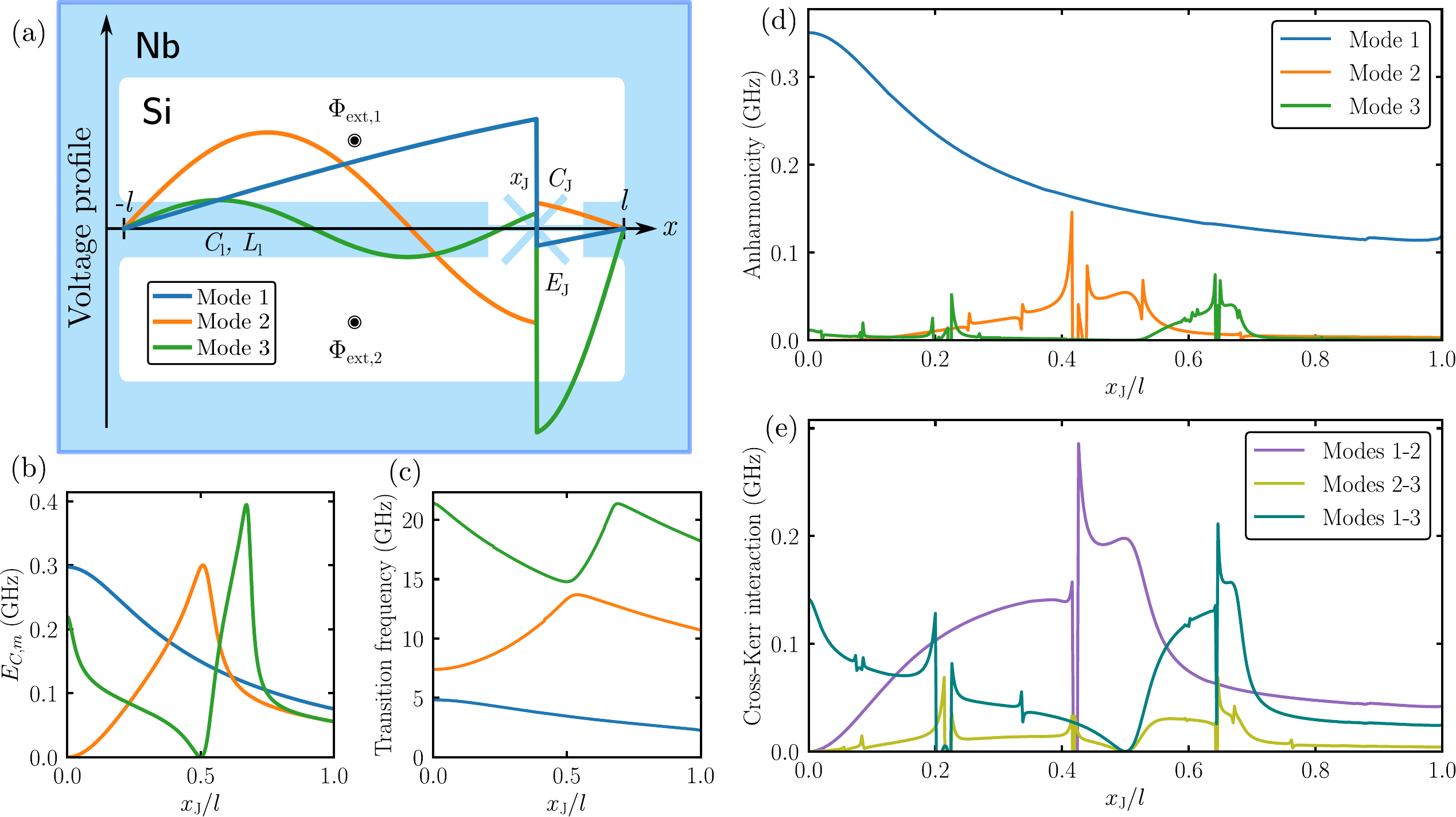}
\caption{\label{fig:schematic_a_d} (a) Schematic representation of a unimon circuit featuring a single Josephson junction ($E_\text{J}$, $C_\text{J}$) that is displaced from the center of the coplanar waveguide (CPW) resonator ($x_\text{J} \neq 0$) with a total length of $2l$. The CPW capacitance and inductance per unit length are denoted as $C_l$ and $L_l$. The white regions symbolize silicon (Si), and the light blue areas represent niobium (Nb), excluding the Josephson junction composed of aluminum. The external magnetic flux biases through the first and second loops are labeled as $\Phi_{\text{ext},1}$ and $\Phi_{\text{ext},2}$. Voltage profiles for the first three modes ($m = 1,2,3$) are illustrated with solid lines. The displacement of the Josephson junction renders its coupling to all modes to be more evenly distributed, as emphasized by the discontinuity in the voltage profile. Please note that the color-coding for the modes remains constant throughout the figure. (b)~The effective charging energy $E_{C,m} = c_m^2 e^2/(2 C_\text{eff})$, (c)~transition frequency $(E_{m,1} - E_{m,0})/h$ of mode $m$, (d)~anharmonicity [see Eq.~\eqref{eq:alpha_m}] of mode $m$, and (e)~cross-Kerr interaction [see Eq.~\eqref{eq:cross_kerr_mn}] between modes $m-n$ as functions of the displacement $x_{\rm J}$ of the Josephson junction.  For panels (b)--(e), the simulations were conducted using the parameter values from Table~\ref{tab:parameter_sets} with the number of low-lying modes set at $M_0 = 8$. The results depicted in panels (c)--(e) are derived from the energy levels featured in Fig.~\ref{fig:energy_level_spagetti}(a). As a consequence, the avoided crossings lead to visible discontinuities in these results.}
\end{figure*}

\section{\label{sec:intro}Introduction}
%{\bf Wallace to write the first version of intro.}
%%%%%%%%%% Introduction %%%%%%%%%%
Superconducting circuits are one of the most promising platforms to enable fault-tolerant quantum computing~\cite{gambetta_building_2017}. However, reaching the stage where such devices become useful in practical applications still seems a major challenge, calling for gate fidelities and coherence times beyond the current state-of-the-art qubits, such as transmons~\cite{koch_charge-insensitive_2007,place_new_2021,wang_transmon_2022}. Even to achieve useful quantum advantage in the on-going noisy-intermediate-scale-quantum (NISQ) era~\cite{preskill_quantum_2018}, gate fidelities exceeding $99.99\%$ for both single-qubit and two-qubit gates may be required, which has not been achieved in superconducting circuits yet. Thus, increasing the quality of superconducting qubits through design and fabrication is one of the greatest on-going technical challenges in the field.

%Significant progress has been made for superconducting qubits since their first measurements dated more than two decades ago~\cite{nakamura_coherent_1999}. Most of this improvement is a result of better fabrication techniques and materials~\cite{place_new_2021,wang_transmon_2021} combined with different qubit designs exploiting several desired features. While transmon qubits~\cite{koch_charge-insensitive_2007} are still the most well-established ones with coherence times reaching up to hundredths of milliseconds~\cite{place_new_2021,wang_transmon_2021}, the particularities of their design seem a natural limiting factor to speed up qubit operations due to the relatively low anharmonicities produced by the transmon circuit.

Recently, different types of unconventional qubits combining desired features have been proposed as alternatives to transmons~\cite{manucharyan_fluxonium_2009,koch_charging_2009,bao_fluxonium_2022,zhang_universal_2021,somoroff_millisecond_2023,ding_high-fidelity_2023,nguyen_blueprint_2022,nguyen_high-coherence_2019,hazard_nanowire_2019,peruzzo_geometric_2021,yan_engineering_2020,ye_engineering_2021,zorin_superconducting_2009,liu_quantum_2023,pechenezhskiy_superconducting_2020,chirolli_quartic_2023,gyenis_experimental_2021,kalashnikov_bifluxon_2020,smith_superconducting_2020,richer_inductively_2017,hassani_inductively_2023}. One of them, the fluxonium, has demonstrated coherence times of the order of milliseconds~\cite{somoroff_millisecond_2023}, and averaged gate fidelities exceeding $99.99\%$ for single-qubit gates~\cite{somoroff_millisecond_2023} and $99.7\%$~for two-qubit gates~\cite{bao_fluxonium_2022, ding_high-fidelity_2023}, thus providing alternative routes towards large-scale quantum computers. However, the involved architecture of the fluxonium may limit its reproducibility in fabrication. In addition, its low frequency requires special techniques for control, readout, and reset~\cite{zhang_universal_2021}. 

The unimon is another unconventional qubit recently proposed and tested experimentally~\cite{hyyppa_unimon_2022}. The unimon circuit exhibits simplicity since it consists of an inductively shunted single Josephson junction that can be biased by external flux. In contrast to the fluxonium, the unimon operates in the regime where the inductive energy of the shunt is mostly cancelled by the quadratic component of the Josephson potential.
%of both large Josephson and inductive energies, $E_{\text{J}}$ and $E_{\text{L}}$, compared to the charging energy $E_{\text{C}}$. 
Consequently, the unimon circuit promotes not only high anharmonicity but also full insensitivity to dc charge noise and reduced sensitivity to homogeneous flux noise. 

Despite the promising first experimental results of the unimon exhibiting 20\% anharmonicities exceeding $0.7$~GHz and  $99.9\%$ single-qubit gate fidelity~\cite{hyyppa_unimon_2022}, there is an urgent demand for a detailed understanding of its underlying physics. In particular, a comprehensive grasp of the fundamental multimode effects originating from the coplanar waveguide (CPW) resonator is important, since theoretical models employed in the description of the unimon to date have been restricted to the first and second-lowest normal modes of the system~\cite{hyyppa_unimon_2022}. Even though they provide a good qualitative agreement with the experimental results, these models do not fully capture the influence of the high-frequency modes. Hence, there is a demand in development of more involved models of the unimon.

In this paper, we develop the theory of multimode unimon circuits and address the physical phenomena induced by its high-frequency modes. Starting from the continuous distributed-element circuit, we quantize the system obtaining an auxiliary-mode Hamiltonian that is equivalent to the effective Hamiltonian obtained using the path-integral-based approach in Ref.~\cite{hyyppa_unimon_2022}.
We then proceed with a partial linearization procedure~\cite{nigg_black-box_2012,solgun_blackbox_2014,bourassa_josephson-junction-embedded_2012,mortensen_normal_2016,smith_quantization_2016,leib_networks_2012} and find a renormalization of the Josephson energy~\cite{leib_networks_2012, leger_observation_2019} that was overlooked in Ref.~\cite{hyyppa_unimon_2022}. The Hamiltonian is then represented in a single-mode unimon basis that allows for efficient numerical diagonalization within the low-energy subspace. We put forward an efficient method to obtain corrections of the qubit energies and anharmonicities induced by the coupling of the qubit mode to a several unoccupied high-frequency modes. For typical unimon parameters, we find anharmonicity reductions of roughly $30\%$ when up to eight modes are taken into account.

Previously, unimon has been studied in the special case where junction is located at the center of the circuit, leading to half of the modes being decoupled from the junction. Importantly, our results indicate that an asymmetric arrangement where the junction is offset from the center provides a rich profile of distributed nonlinearity, which is expressed by significant self- and cross-Kerr couplings between the modes. Particularly, the numerical findings from both two- and three-mode scenarios display moderate nonlinearity across all modes concurrently, hinting at the potential for multimon-like qubit operations~\cite{roy_multimode_2018,roy_programmable_2020,roy_implementation_2017} within unimon circuits. However, to achieve high-fidelity multiqubit operations, unimon circuits with more intricate designs or additional components are likely needed. We further study the accuracy of our numerical findings by analytically solving for self- and cross-Kerr interactions within the harmonic-oscillator basis.

This paper is organised as follows. In Sec.~\ref{sec:model}, we introduce our model and notation, detailing the linearization procedure and obtaining spectrum of the qubit. In Sec.~\ref{sec:single-mode}, we describe the effects of the high-frequency modes on the unimon-qubit mode comparing them with the results predicted by the single-mode approximation and with the auxiliary-mode models introduced in Ref.~\cite{hyyppa_unimon_2022}. In Sec.~\ref{sec:multimode}, we study the multimode structure of the unimon, focusing on the self and cross-Kerr terms. Our conclusions are presented in Sec.~\ref{sec:conclusions}.

\section{\label{sec:model}Multimode model for the unimon}

We study a single-junction unimon circuit with multiple modes taken into account. The system is schematically illustrated in Fig.~\ref{fig:schematic_a_d}(a). Our primary motivation lies in discerning how the position of the Josephson junction within the half-wavelength resonator %, denoted as $x_\text{J} \in \bigl[-l, l \bigr]$, 
influences the nonlinearity of the modes. Key indicators of this nonlinearity, namely the mode effective charging energy, transition frequency, anharmonicity, and cross-Kerr coupling between modes, are presented in Fig.~\ref{fig:schematic_a_d}(b)--(e). To accurately compute these values, we introduce an effective theoretical model, which is elaborated upon in the subsequent sections.

\subsection{Effective multimode unimon Hamiltonian} \label{sec:Effective multimode unimon Hamiltonian}

As a starting point for our multimode treatment of the unimon circuit, we use a Hamiltonian that comprises a nonlinear mode with $\hat{\Psi}$ describing the magnetic flux difference across the Josephson junction and being conjugate to the charge $\hat{Q}$, and $M$ included linear auxiliary modes with fluxes $\hat{\chi}_m$ that are conjugate to charges $\hat{\Xi}_m$, where $m = 1,...,M$. The relevant nonzero single-operator commutation relations these operators satisfy are $[\hat{\Psi}, \hat{Q}] = \ii \hbar$ and $[\hat{\chi}_k, \hat{\Xi}_m] = \ii \hbar \delta_{km}$. A detailed derivation for the Hamiltonian is provided in the Appendix~\ref{app:Derivation}, and hence, we begin our treatment from the auxiliary-mode Hamiltonian
\begin{align}
    \hat{H}_\text{aux} = &\frac{\hat{Q}^2}{2 C_\text{eff}} + \frac{1}{4 l L_l} \Bigl( \hat \Psi + \Phi_\text{diff} - \phi_0\Bigr)^2 \notag \\
    & - E_{\text{J}} \cos \biggl[ \frac{2 \pi}{\Phi_0} ( \hat{\Psi} - \phi_0 ) \biggr] \label{eq:H_aux} \\
    &+ \sum_{m = 1}^M \Bigg[ \frac{\hat{\Xi}_m^2}{2 C_\text{eff}} + \frac{1}{2}C_\text{eff}\Omega_m^2 \biggl( \hat{\chi}_m + \frac{\xi_m \hat \Psi}{C_\text{eff}\Omega_m^2} \biggr)^2 \Bigg], \notag
\end{align}
%\textcolor{red}{
%\begin{multline*}
%    \frac{1}{2} C_\mathrm{eff} \Omega_m^2 \hat \chi_m^2 + \xi_m \hat \chi_m \hat \Psi \\ = 
%    \frac{1}{2} C_\mathrm{eff} \Omega_m^2 \left(\hat \chi_m + \frac{\xi_m \hat \Psi}{C_\mathrm{eff} \Omega_m^2}\right)^2 - \frac{1}{2} \frac{\xi_m^2 \hat \Psi^2}{C_\mathrm{eff} \Omega_m^2}
%\end{multline*}
%\begin{multline*}
%    \frac{\hat \Psi^2}{2 L_\mathrm{eff}} - \frac{1}{2} \sum\limits_{m=1}^M \frac{\xi_m^2 \hat \Psi^2}{C_\mathrm{eff} \Omega_m^2} = \\
%    \frac{\hat \Psi^2}{2}\left\{
%    \frac{1}{2l L_l} + \sum\limits_{m=1}^M
%    \left[ 
%    \frac{\cos^2\left(\Omega_m x_\mathrm J / v + m \pi / 2\right)}{l L_l}
%    \right. \right . \\ 
%    \left .
%    \left.
%    -\frac{C_\mathrm{eff} \Omega_m^2}{l L_l C_\mathrm{eff} \Omega_m^2} \cos^2\left(\Omega_m x_\mathrm J / v + m\pi / 2\right)
%    \right]
%    \right\} = \frac{\hat \Psi^2}{4 l L_l}
%end{multline*}
%}
where $C_{\text{eff}}$ is the effective capacitance, $L_l$ is the resonator inductance per unit length, $2l$ is the resonator length, $\Phi_{\text{diff}} = (\Phi_{\text{ext},1} - \Phi_{\text{ext},2})/2$ is the half difference of the external magnetic fluxes shown in Fig.~\ref{fig:schematic_a_d}(a), $\phi_0$ is the dc magnetic flux offset across the Josephson junction, $E_\text{J}$ is the Josephson energy,  $\Phi_0$ denotes the flux quantum, $\Omega_m$ is the resonance angular frequency of the auxiliary mode $m$, and $\{\xi_m\}$ are the coupling strengths of the auxiliary modes to the nonlinear mode. The definition for the effective lumped-element capacitance is
\begin{align}
    C_{\text{eff}} &= C_{\text{J}} + \frac{C_l (l^2 + 3 x_\text{J}^2)}{6l} - \sum_{m = 1}^M \frac{\cos^2 \bigl( \frac{\Omega_m x_\text{J}}{v} + \frac{m \pi}{2} \bigr)}{\Omega_m^2l L_l}, \label{eq:C_eff}
\end{align}
and the auxiliary-mode coupling strengths assume the form
\begin{equation} \label{eq:xi_m}
\begin{split}
    \xi_m = \sqrt{\frac{C_\text{eff}}{l L_l}} \Omega_m \cos \biggl( \frac{\Omega_m x_\text{J}}{v} + \frac{m \pi}{2} \biggr),
\end{split}
\end{equation}
where $v = 1/\sqrt{L_l C_l}$ is the phase velocity, $C_\text{J}$ is the capacitance of the Josephson junction, $C_l$ denotes the capacitance per unit length, and $x_\text{J} \in [-l,l]$ represents the location of the Josephson junction in the CPW resonator.

The auxiliary-mode Hamiltonian in Eq.~\eqref{eq:H_aux} contains $M$ auxiliary modes. Although this expression becomes exact as $M \rightarrow \infty$, numerical calculations necessitate limiting the number of modes at a finite $M$. How to choose $M$ is a relevant question for our multimode model and is addressed in more detail in Sec.~\ref{sec:single-mode}. In addition, we note that the version of the auxiliary-mode Hamiltonian introduced here differs slightly from the one presented in Ref.~\cite{hyyppa_unimon_2022}. The distinction lies in the dc flux offset parameter $\phi_0$ that we introduce (Appendix~\ref{app:Derivation}) to facilitate our derivation of the multimode model.

The auxiliary modes couple to the nonlinear mode inductively, providing corrections to the unimon qubit energy levels. The auxiliary-mode Hamiltonian in Eq.~\eqref{eq:H_aux} is well-adapted for studying the unimon circuit if our interest is focused only on the lowest mode, as in Ref.~\cite{hyyppa_unimon_2022}. Since the auxiliary-mode frequencies are integer multiples of the lowest auxiliary mode, $\Omega_m = \pi m v / (2l) $, only the few lowest modes are energetically close enough to the nonlinear mode $\hat{\Psi}$ to significantly interact with it. Furthermore, the location of the Josephson junction can be chosen strategically to leave certain auxiliary modes uncoupled, thus easing the load for numerical calculations. In an optimal scenario, this approach allows for fairly accurate results for unimon qubit energy levels to be obtained by incorporating just a single auxiliary mode, effectively reducing the problem to solving a two-dimensional Schrödinger equation in the flux basis~\cite{hyyppa_unimon_2022}. However, extending the consideration from the energy levels of the lowest mode to those of a multimode system necessitates an alternative approach, primarily due to the rapid escalation of computational demand as more auxiliary modes are included.

To adapt the Hamiltonian of the unimon circuit for numerical analysis in cases involving multiple modes, we divide the auxiliary-mode Hamiltonian in Eq.~\eqref{eq:H_aux} into linear and nonlinear parts, $\hat{H}_\text{aux} = \hat{H}_\text{lin} + \hat{H}_\text{nl}$, by expanding the nonlinear Josephson term and moving the resulting quadratic term to the linear part (Appendix~\ref{app:Linearization}). To elucidate different modes in the system and simplify subsequent analysis, we find the classical normal modes of the linear part, $\hat{H}_{\text{lin}}$, using a basis transformation. This process, which effectively removes the linear coupling between the flux operators in Eq.~\eqref{eq:H_aux}, is detailed in the Appendix~\ref{app:Linearization}. We note that similar linearization procedures to find the normal modes of system have been employed in earlier works, as seen in Refs.~\cite{nigg_black-box_2012,solgun_blackbox_2014,bourassa_josephson-junction-embedded_2012,mortensen_normal_2016,smith_quantization_2016,leib_networks_2012}.

The basis transformation gives rise to the normal-mode flux operators and the corresponding conjugate charge operators which are denoted by $\hat{\phi}_m$ and $\hat{q}_m$, respectively. These new operators continue to satisfy the canonical commutation relations $[\hat{\phi}_k, \hat{q}_m] =\ii\hbar \delta_{km}$. The transformation also yields the normal-mode representation of the magnetic flux across the Josephson junction:
\begin{equation} \label{eq:Psi_transformed}
\hat{\Psi} = \sum_{m = 1}^{M+1}c_m \hat{\phi}_m,
\end{equation}
where the constant factors $c_m$ are coefficients determined by the transformation and describe the contribution from each normal mode to the overall magnetic flux. In addition, the diagonalization process uncovers the normal-mode frequencies, denoted by $\omega_m/(2\pi), \; m = 1,...,M+1$. Insertion of Eq.~\eqref{eq:Psi_transformed} into the nonlinear part $\hat H_\text{nl}$ gives us the full normal-mode representation of the auxiliary-mode Hamiltonian as
\begin{equation} \label{eq:H_nm}
\begin{split}
    \hat{H}_{\text{aux}} = &\sum_{m = 1}^{M+1} \biggl[ \frac{\hat{q}_m^2}{2 C_\text{eff}} + \frac{\hat{\phi}_m^2}{2 \tilde{L}_m} + I_\text{c} \sin \biggl( \frac{2 \pi \phi_0}{\Phi_0} \biggr) c_m \hat{\phi}_m \biggr] \\
    &- E_{\text{J}} \cos \Biggl[ \frac{2 \pi}{\Phi_0} \Biggl( \sum_{m = 1}^{M+1} c_m \hat{\phi}_m - \phi_0 \Biggr) \Biggr] \\
    &- \frac{1}{2 L_\text{J}} \cos \biggl( \frac{2 \pi \phi_0}{\Phi_0} \biggr) \sum_{\substack{m,k=1 \\ m \neq k}}^{M+1} c_m c_k \hat{\phi}_m \hat{\phi}_k,
\end{split}
\end{equation}
where we have defined the Josephson inductance $L_\text{J}~=~(2 \pi / \Phi_0)^2/E_\text{J}$, the critical current $I_\text{c} = 2 \pi E_\text{J} / \Phi_0 $ and 
\begin{equation} \label{eq:L_eff_tilde}
\tilde{L}_m = \left[ C_\text{eff} \omega_m^2 - \frac{\cos(2 \pi \phi_0 / \Phi_0)}{L_\text{J}}c_m^2 \right]^{-1},
\end{equation}
to represent the effective inductance of the $m$:th normal mode.

Above, we described the Hamiltonian of the unimon circuit in the normal-mode basis of the linearized version of the circuit. However, a challenge still remains, since finding the energy levels of the unimon qubit, while including the interactions from the higher number of modes, requires solving a high-dimensional Schrödinger equation, and thus, an improved basis to find the solution is needed.

In order to simplify the notation, we introduce dimensionless operators $\hat{n}_m = (\hat{q}_m/c_m)/(2e)$ and $\hat{\varphi}_m = 2 \pi c_m \hat{\phi}_m / \Phi_0$, where $e$ is the elementary charge. We divide the Hamiltonian in Eq.~\eqref{eq:H_nm} into parts describing the single-mode unimon Hamiltonians $\hat{H}_{m}$ and the interaction part $\hat{H}_{\text{int}}$, leading to
\begin{equation} \label{eq:Hm_plus_Hint}
\hat{H}_{\text{aux}} = \sum_{m = 1}^{M+1}\hat{H}_{m} + \hat{H}_{\text{int}}.
\end{equation}
These constituent Hamiltonians can be expressed as
\begin{equation} \label{eq:Hm}
\begin{split}
\hat{H}_{m} =  & \ 4 E_{C,m} (\varphi_0 ) \hat{n}_m^2 + \frac{1}{2}E_{L,m}(\varphi_0 ) \hat{\varphi}_m^2 \\
&+  E_\text{J} \bigl[ \sin \bigl(\varphi_0 \bigr) \hat{\varphi}_m - \cos \bigl(\hat{\varphi}_m - \varphi_0 \bigr) \bigr],
\end{split}
\end{equation}
and
\begin{align}
    \hat{H}_{\text{int}} = E_{\text{J}} \Biggl[ & \sum_{m = 1}^{M+1} \cos \bigl(\hat{\varphi}_m - \varphi_0 \bigr) - \cos \Biggl( \sum_{m = 1}^{M+1} \hat{\varphi}_m - \varphi_0 \Biggr) \notag\\
    &- \frac{\cos ( \varphi_0 )}{2} \sum_{\substack{m,k=1 \\ m \neq k}}^{M+1} \hat{\varphi}_m \hat{\varphi}_k \Biggl], \label{eq:H_int} 
    \end{align}
where we have defined an effective charging energy of the $m$:th mode as $E_{C,m} (\varphi_0 ) = c_m^2(\varphi_0 ) e^2/(2 C_\text{eff})$, effective inductive energy as $E_{L,m} (\varphi_0 ) = \Phi_0^2/(2 \pi)^2/ [\tilde{L}_m c_m^2(\varphi_0 )]$, and the dc phase across the junction is denoted as $\varphi_0 = 2 \pi \phi_0 / \Phi_0$. Note that the coefficients $c_m$ introduce the dependence of $E_{C,m}$ and $E_{L,m}$ on $\varphi_0$.

This normal-mode representation of the auxiliary-mode Hamiltonian is beneficial for several reasons. On one hand, it fully separates the single-mode components from the interaction terms. This becomes particularly clear on the first row of Eq.~\eqref{eq:H_int}, where the first term cancels the concealed single-mode terms in the second term (see Appendix~\ref{app:division_to_Hm_and_H_int}). On the other hand, the single-mode unimon Hamiltonian $\hat{H}_m$ can be diagonalized efficiently using the one-dimensional Schrödinger equation, $\hat{H}_m \ket{j_m} = E_{m,j} \ket{j_m}$, where $\ket{j_m}$ and $E_{m,j}$ denote the $j$:th eigenstate of the $m$:th mode and the corresponding eigenenergy. Moreover, the eigenstates of $\hat{H}_m$ form a set of basis states that can be used to represent the interaction part $\hat{H}_\text{int}$ in a matrix form, that subsequently allows an efficient diagonalization of the Hamiltonian in Eq.~\eqref{eq:Hm_plus_Hint} for the purposes of studying the effect of the other modes on the energy levels of the unimon qubit. In practice, this is accomplished by applying the generalized trigonometric sum relation
\begin{equation} \label{eq:cos_infinite_sum}
\begin{split}
    \cos \Biggl( \sum_{i = 1}^N \hat{\varphi}_i \Biggr) = &\sum_{\substack{k = 0 \\ \text{even } k}} ^{N} (-1)^{\frac{k}{2}}
    \sum_{\substack{A\subseteq\{1,...,N \} \\
                  \abs{A} = k}} \\
                  & \times \Bigg[ \prod_{i \in A} \sin \bigl( \hat{\varphi}_i  \bigr) \prod_{i \notin A} \cos \bigl( \hat{\varphi}_i  \bigr) \Bigg]
\end{split}
\end{equation}
%\end{subequations}
for a cosine of sums, and subsequently calculating matrix elements of three different types, $\mel**{j_m}{\cos \bigl( \hat{\varphi}_m \bigr) }{j_m'}$, $\mel**{j_m}{\sin \bigl(\hat{\varphi}_m \bigr) }{j_m'}$, and  $\mel**{j_m}{\hat{\varphi}_m}{j_m'}$, in Eq.~\eqref{eq:H_int}.

To avoid any possible confusion, we emphasize that the last summation in Eq.~\eqref{eq:cos_infinite_sum} represents a sum over all possible subsets of $\{1,...,N\}$ with a size of $k$. The product over $i \notin A$ refers to all elements in $\{1,...,N\}$ not included in $A$. When $k = 0$, $A$ is an empty set, and the product over $i \notin A$ encompasses all elements of $\{1,...,N\}$.

\subsection{Energy cutoff for Hilbert space}

After establishing a matrix representation in the single-mode unimon basis, we need to manage the dimensions of the Hilbert space before proceeding with an efficient diagonalization of the matrix. To this end, we restrict the total size of the Hilbert space, which ensures computational feasibility and accuracy of our numerical approximations.

Although the number of energy levels for each mode is theoretically infinite, ideal quantum computation takes place in a finite-dimensional space and, therefore, is compatible with the concept of an energy cutoff $E_{\text{cutoff}}$. Consequently, we disregard all eigenstates of the single-mode Hamiltonian $\hat{H}_m$ with energies exceeding $E_{\text{cutoff}}$. The value of the energy cutoff is determined by the convergence of the low-energy eigenstates of the full Hamiltonian that we aim to accurately model. Since the bare frequencies of the modes increases with the mode number $m$, we only need to consider states beyond the ground state for a limited number of modes. This is attributed to the fact that the energy required to excite such a mode exceeds the established energy cutoff.

We categorize the complete set of modes ($m={ 1,2,...,M+1 }$) into two distinct groups, referred to as the \emph{lower modes} and \emph{higher modes}. The lower modes, defined by an integer $M_0$ and $m\in\{ 1,...,M_0 \}$, include all modes where any excited states are considered. Conversely, for the higher modes, for which $m\in{ M_0 + 1,...,M+1 }$, only their vacuum state is included due to the energy cutoff.

To simplify numerical computations, we assume that the system is operated at a flux sweet spot where $\Phi_\text{diff}~=~\Phi_0/2$ and $\varphi_0~=~\pi$. This condition results in a symmetry $\bra{\varphi_m}\ket{j_m}=|\bra{-\varphi_m}\ket{j_m}|$. For the vacuum state, we have $\bra{\varphi_m}\ket{0_m}=\bra{-\varphi_m}\ket{0_m}$, and hence the expectation values of operators anti-symmetric in $\hat{\varphi}_m$ vanish for the vacuum sate. For example,  $\mel**{0_m}{\sin \bigl(\hat{\varphi}_m \bigr) }{0_m}=0$ and $\mel**{0_m}{\hat{\varphi}_m}{0_m}=0$. Interestingly, this implies that the impact of the higher modes on the system primarily contributes to a renormalization of the Josephson energy, denoted as
\begin{equation} \label{eq:E_J_renorm}
    \tilde{E}_\text{J} = E_\text{J} \prod_{m = M_0 + 1}^{M+1} \mel**{0_m}{\cos \bigl( \hat{\varphi}_m \bigr) }{0_m}.
\end{equation}
A more comprehensive exploration of this renormalization effect and its implications can be found in Sec.~\ref{sec:single-mode}.

After incorporating all of the above-described steps, we arrive at the final form of the total Hamiltonian
\begin{widetext}
\begin{equation} \label{eq:H_nm_final}
    \hat{\tilde H}_{\text{aux}} = \sum_{m = 1}^{M_0} \hat{H}_m -E_{\text{J}} \Biggl[ \sum_{n = 1}^{M_0} \cos \bigl(\hat{\varphi}_n \bigr) - \frac{1}{2} \sum_{\substack{l,k=1 \\ l \neq k}}^{M_0} \hat{\varphi}_l \hat{\varphi}_k \Biggl]
    + \sum_{\substack{j = 0 \\ \text{even } j}} ^{M_0} (-1)^{\frac{j}{2}}
    \sum_{\substack{A\subseteq\{1,...,M_0 \} \\
                  \abs{A} = j}}
                   \tilde{E}_\text{J}\Biggl[ \prod_{i \in A} \sin \bigl( \hat{\varphi}_i  \bigr) \prod_{i \notin A} \cos \bigl( \hat{\varphi}_i  \bigr) \Biggr],
\end{equation}
\end{widetext}
where we have used Eq.~\eqref{eq:cos_infinite_sum} and truncated the Hilbert space based on the energy cutoff. We consider this expression to be one of the main results of this paper.

In our model, the Hamiltonian described by Eq.~\eqref{eq:H_nm_final} is expressed in a matrix form using the single-mode unimon basis, where matrix elements are expressed as
\begin{equation} \label{eq:H_nm_matrix}
\big[\tilde H_\text{aux}\big]_{i_0 j_1 ... k_{M_0}}^{i_0' j_1' ... k_{M_0}'}~=~\mel{i_0 j_1 ... k_{M_0}}{\hat{\tilde H}_\text{aux}}{i_0' j_1' ... k'_{M_0}},
\end{equation}
and each state $\ket{i_0 j_1 ... k_{M_0}}$ must meet the energy cutoff condition
\begin{equation} \label{eq:cutoff_condition}
\mel**{i_0 j_1 ... k_{M_0}}{\sum_{m = 1}^{M_0} \hat{H}_{m}}{i_0 j_1 ... k_{M_0}}~\leq~E_\text{cutoff}.
\end{equation}
By carefully selecting the energy cutoff, the diagonalization of the matrix can be accomplished with adequate numerical efficiency and accuracy.

\begin{figure}[t]
\includegraphics[scale=0.5]{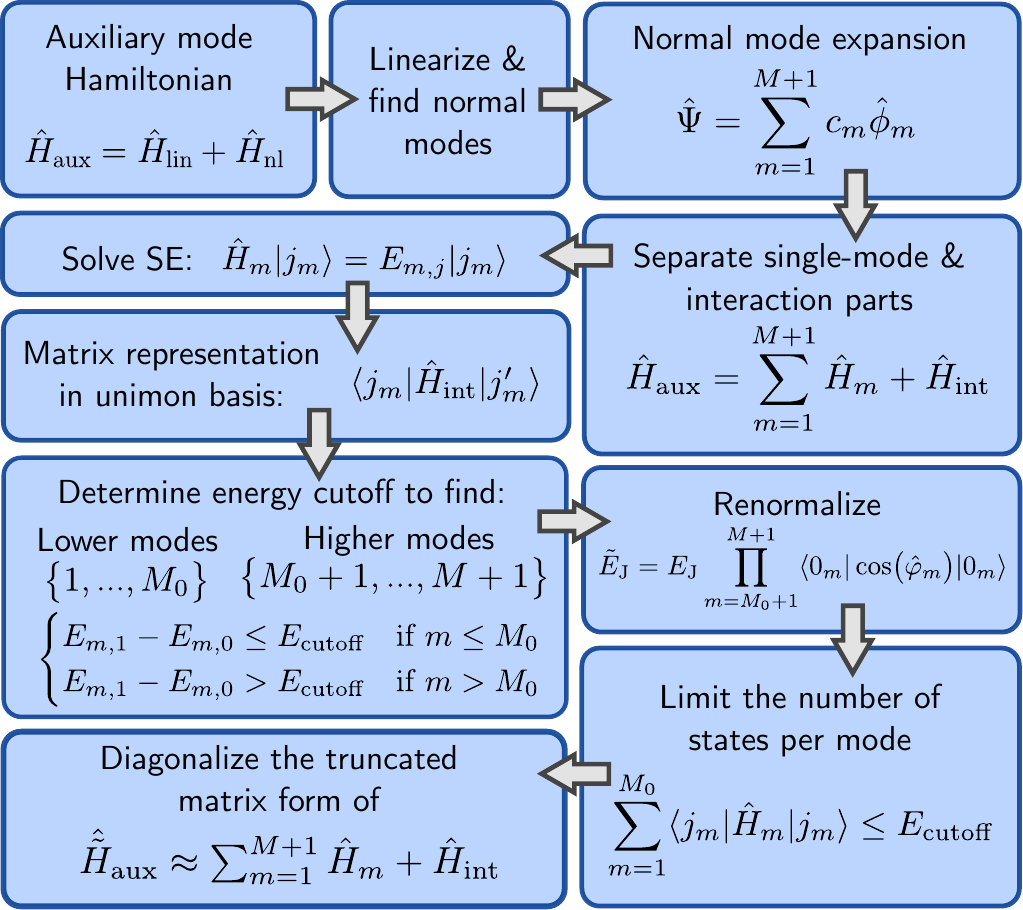}
\caption{\label{fig:model_process_chart} Process chart illustrating the method for solving the energy levels and eigenstates of the unimon circuit in a multimode scenario. This schematic provides a visual summary of the more detailed instructions and definitions given in Sec.~\ref{sec:model}.% Initiating with the first step, we identify the normal modes of the linear part of $\hat{H}_\text{aux}$. Moving to the second step, we segregate the full Hamiltonian into single-mode sections and their corresponding interactions, followed by solving these single-mode parts. For the third step, we utilize the resultant eigenstates as a new basis for the interaction part. In the fourth step, we define an energy cutoff value, derived from the single-mode energy levels, that distinguishes between lower and higher modes. During the fifth step, we compute the renormalization coefficient, taking into account the influence of the higher modes. In the sixth step, we apply the energy cutoff value to limit the number of states in the lower modes. This prepares us to diagonalize the full matrix, which is the final procedure of our process.
}
\end{figure}

The above-introduced process of solving the eigenstates of the unimon circuit is visually summarized in Fig.~\ref{fig:model_process_chart}.

\subsection{Labeling of eigenstates} \label{subsec:Labeling of eigenstates}

\begin{figure}[t]
\includegraphics[scale=0.50]{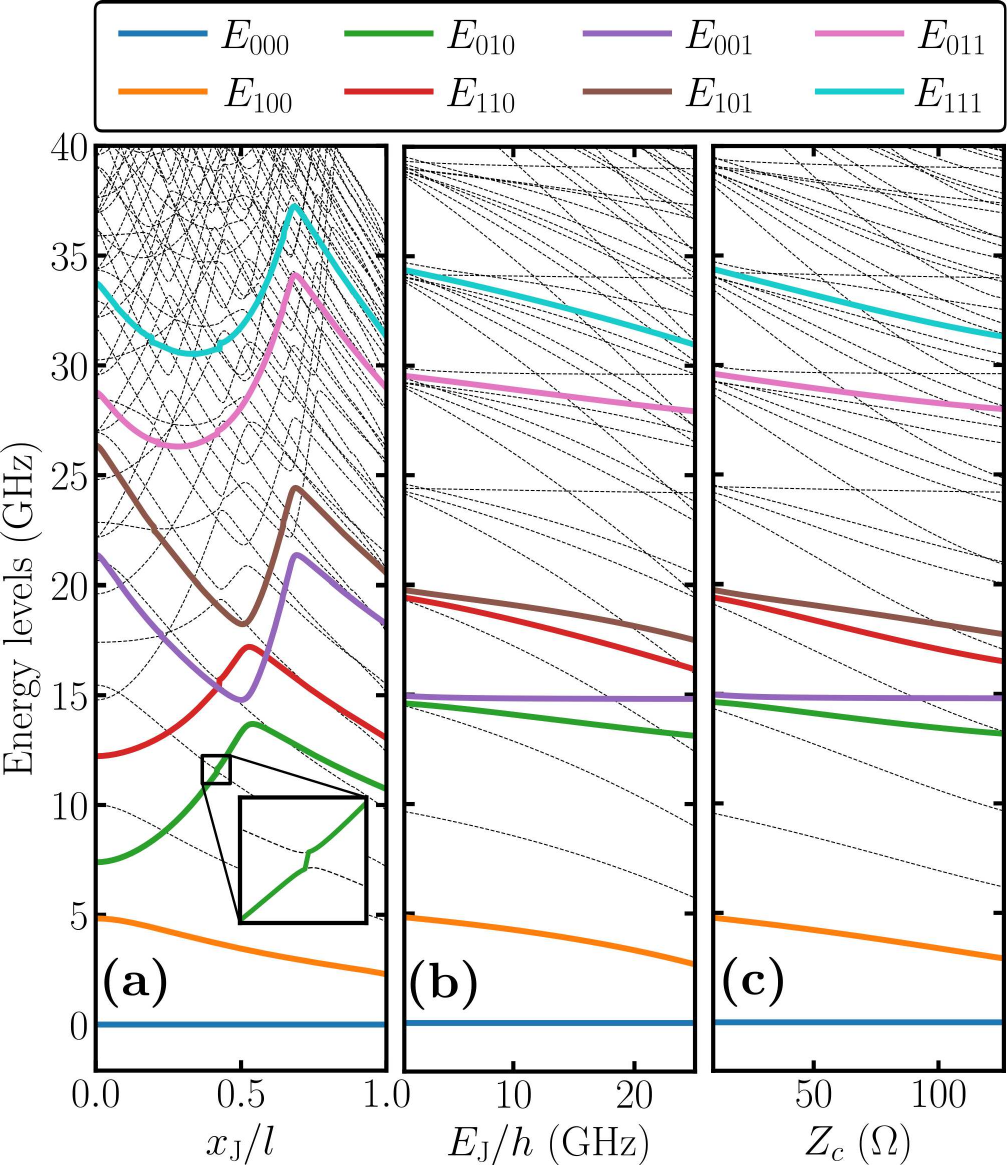}
\caption{\label{fig:energy_level_spagetti} Energy levels of the unimon circuit as functions of (a)~the location of the Josephson junction $x_\text{J}$, (b)~Josephson energy $E_\text{J}$, and (c)~characteristic impedance $Z_\text{c}$ of the CPW. The solid lines represent the qubit subspace of the three first modes, for which the rest of the modes are in the vacuum state. All other states are represented by dashed lines. The inset of panel (a) provides a close view of an example avoided crossing. For these simulations, we used the parameters from Table~\ref{tab:parameter_sets}, with the number of lower modes set to $M_0 = 8$, except that the sweep of the characteristic impedance is conducted such that the frequency of the lowest auxiliary mode, given by $\Omega_1 = \pi / (2l \sqrt{L_l C_l})$, remains constant.}
\end{figure}

Upon diagonalizing the matrix, we acquire a new set of eigenstates along with corresponding energy eigenvalues, which can be interpreted as perturbed versions of the single-mode unimon eigenstates. To enhance our understanding of the effects of the interactions between the modes, it is fruitful to study the energy differences between the perturbed and non-interacting scenarios.

Although visual inspection of the energy levels can yield insight in specific cases, this approach tends to become increasingly demanding in general. To streamline this process, we compare the energy eigenstates of the interacting Hamiltonian with those of the non-interacting case. In practice, this is achieved by calculating the state overlaps $\abs{\braket{i_0 j_1 ... k_{M_0} }{\alpha_n}}$, where $\ket{\alpha_n}$ represents the $n$:th eigenstate of $\tilde H_\text{aux}$ and $\ket{i_0 j_1 ... k_{M_0}}$ corresponds to an eigenstate of the non-interacting Hamiltonian $\sum_{m = 1}^{M_0}\hat{H}_{m}$. We then identify the state $\ket{i_0 j_1 ... k_{M_0}}$ that exhibits the maximum overlap with $\ket{\alpha_n}$, and label $\ket{\alpha_n}$ as the perturbed counterpart of $\ket{i_0 j_1 ... k_{M_0}}$.

Note that our labeling method may produce ambiguous results. This ambiguity arises from transverse-type interactions introduced by the interaction term, leading to strong hybridization between eigenstates near suitable degeneracy points. %In our research, the effects of hybridization are visually apparent in graphs illustrating the behavior of the energy levels or derived quantities such as anharmonicity.
Figure~\ref{fig:energy_level_spagetti}(a) illustrates the effects of the hybridization in the energy level diagram, exhibiting avoided crossings between levels. Where eigenstate labeling is applied on quantities such as anharmonicity or cross-Kerr interaction, the effects of hybridization are exposed through sudden discontinuities as found in Fig.~\ref{fig:schematic_a_d}(d)--(e).

\section{\label{sec:single-mode}Multimode effects in the unimon circuit}

Our next step is to analyze how the modes beyond the lowest mode affect the energy levels and anharmonicity of the unimon qubit. First, we focus on the anharmonicity calculated by using the renormalization model. In contrast to the previous single-mode approximations, as shown in Eq.~\eqref{eq:Hm}, here we use the renormalized Josephson energy, denoted as $\tilde{E}_\text{J}$. In addition, we discuss the anharmonicity results from the multimode model detailed in Sec.~\ref{sec:model} and juxtapose them with the outcomes of the renormalization model. As a further point of comparison, we also discuss the distinctions between these findings and those acquired using the auxiliary-mode model of Ref.~\cite{hyyppa_unimon_2022}.

\begin{table} [t]
\caption{\label{tab:parameter_sets}
Physical parameters used in the simulations of this paper unless otherwise explicitly stated.}
\begin{ruledtabular}
\begin{tabular}{cccccccc}
 \shortstack{$2l$ \\ (mm)} & \shortstack{$L_l$ \\ ($\upmu$H/m)} & \shortstack{$C_l$ \\ (pF/m)}& \shortstack{$E_\text{J} / h$ \\ (GHz)} & \shortstack{$x_\text{J} / l$ \\ (-)} & \shortstack{$Z_{\text{c}}$ \\ ($\Omega$)} & \shortstack{$\Omega_1 / (2 \pi)$ \\ (GHz)} & \shortstack{$\Phi_\text{diff}/\Phi_0$ \\ (-)}\\
\hline
8.0 & 0.821 & 87.1 & 19.0 & 0.51 & 97.1 & 7.39 & 0.5 \\
\end{tabular}
\end{ruledtabular}
\end{table}

\subsection{Effect of renormalization}

\begin{figure*}[t]
\includegraphics[scale=0.48]{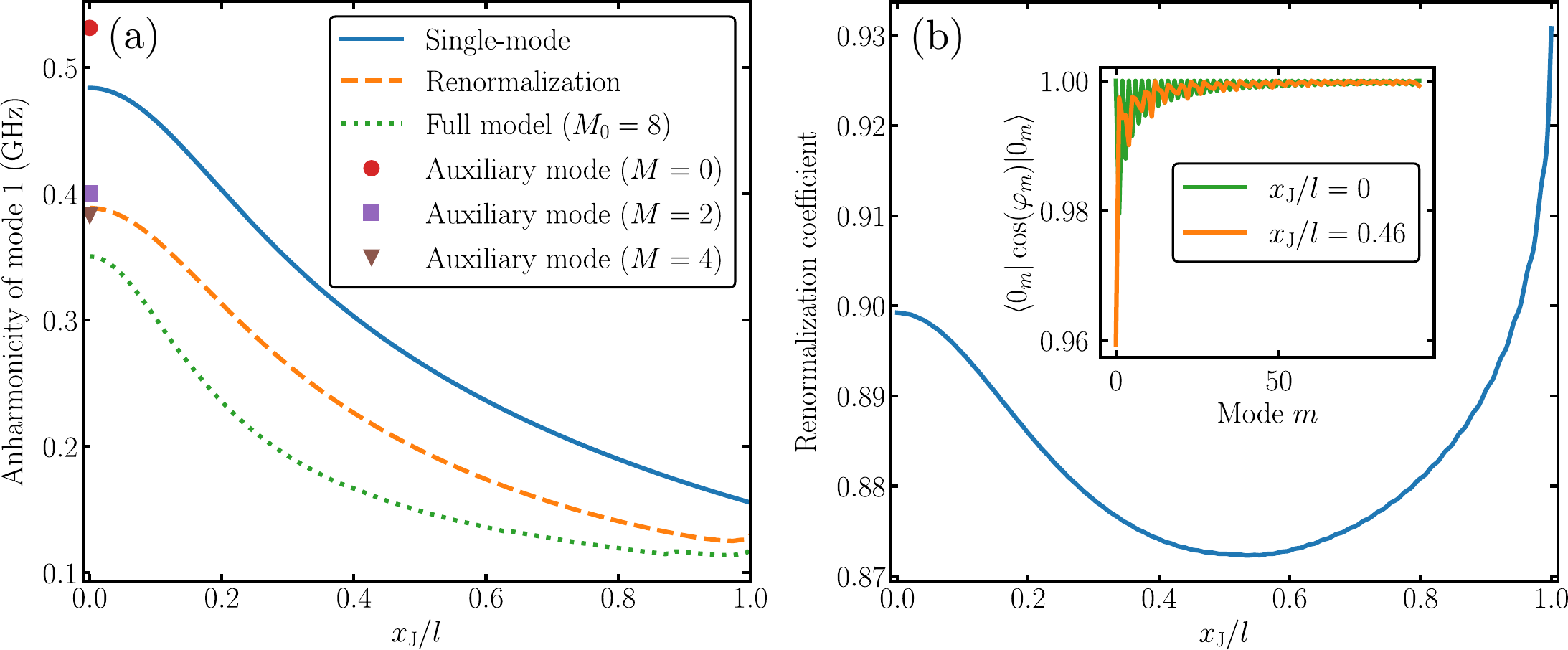}
\caption{\label{fig:mode1_anharmonicity} (a)~Anharmonicity of the unimon qubit and (b)~the renormalization coefficient $\tilde{E}_\text{J}/E_\text{J}$ as function of the location of the junction $x_{\rm J}$. We show in (a) the anharmonicity obtained using the single-mode approximation (solid line), renormalization model (dashed line), and the multimode model (dotted line) detailed in Sec.~\ref{sec:model}. The anharmonicities computed directly from the auxiliary-mode Hamiltonian at $x_\text{J} = 0$ are also indicated by markers, accounting for different numbers of auxiliary modes. The inset in (b) shows the contribution from each mode to the renormalization coefficient for $x_\text{J}/l = 0$ (green line) and $x_\text{J}/l = 0.46$ (orange line) [see Eq.~\eqref{eq:E_J_renorm}]. These numerical calculations are carried out using the parameters listed in Table~\ref{tab:parameter_sets}.}
\end{figure*}

As indicated in Eq.~\eqref{eq:E_J_renorm}, at the sweet spot $\varphi_0 = \pi $, the higher modes influence the energy levels of the unimon even in their vacuum state, where these modes still exhibit zero-point phase fluctuations. These fluctuations couple to the unimon mode, effectively causing a renormalization of the Josephson energy within the system~\cite{leib_networks_2012, leger_observation_2019}. Given that the renormalization coefficient of Eq.~\eqref{eq:E_J_renorm} is simply a product of the vacuum state expectation values of the cosine function, it follows that $\tilde{E}_\text{J}/E_\text{J} \leq 1$. As a result of the renormalization, we thus observe a decrease in the anharmonicity of the unimon qubit.

Interestingly, this decrease in anharmonicity due to the renormalization of the Josephson energy offers an explanation for the discrepancies observed between the two models used in Ref.~\cite{hyyppa_unimon_2022}: the single-mode approximation and the auxiliary-mode model based on path integrals. The auxiliary-mode model in Eq.~\eqref{eq:H_aux}, which includes the nonlinear mode and a few lowest linear modes, is expected to provide a more accurate results than the single-mode approximation model, which only incorporates one normal mode [Eq.~\eqref{eq:Hm}].

In Fig.~\ref{fig:mode1_anharmonicity}(a), we display the discrepancy in qubit anharmonicities. Between the single-mode and auxiliary-mode ($M = 2$) approximations, the difference is roughly 20\% at $x_{\rm J}=0$. By including the renormalization, we observe an approximate 20\% decrease in the anharmonicity compared to the single-mode approximation, bringing the result appreciably close to the predictions of the auxiliary-mode approach. In Fig.~\ref{fig:mode1_anharmonicity}(b), we further detail the behavior of the renormalization coefficient, which changes only weakly as a function of $x_\text{J}$.

It is worth mentioning that based on physically motivated mode-cutoff frequncies~\cite{gely_convergence_2017, leib_networks_2012}, the number of modes $M$ included in the renormalization process should be finite. Here we introduce a cutoff that is based on the magnitude of the superconductor gap parameter $\Delta_\text{gap}$. Namely, we set $\Omega_m \leq 2 \Delta_\text{gap}$, which imposes a limit on the number of modes~\cite{leib_networks_2012,filipp_multimode_2011,frisk_kockum_designing_2014}. However, the renormalization coefficient appears to be insensitive to the precise number of modes included and seems to converge as $M \rightarrow \infty$. This is evidenced by the inset of Fig.~\ref{fig:mode1_anharmonicity}(b), where the contribution from each mode around the cutoff frequency ($M \approx 100$) is negligible. More detailed discussion regarding the convergence of the renormalization coefficient is found in Appendix~\ref{app:scaling}.

Our numerical findings shown in Fig.~\ref{fig:mode1_anharmonicity} demonstrate that the renormalization of the Josephson energy leads to anharmonicities close to those obtained from the flux basis solution of the auxiliary-mode Hamiltonian. Importantly, compared to the single-mode-approximation model, the renormalization model achieves a comparable level of accuracy to the auxiliary-mode approach ($M = 2$), with virtually no increase in computational load.

\subsection{Effects beyond renormalization}

In addition to the renormalization effect of vacuum states, we aim to understand how the energy levels of the unimon qubit are influenced when excited eigenstates in modes $m >1$ are included within the computational Hilbert space after truncation. Therefore, we employ the comprehensive multimode method to solve for the energy levels and anharmonicities of the full Hamiltonian in the single-mode basis as shown in Eq.~\eqref{eq:H_nm_matrix}. Our findings show that the anharmonicity decreases further in comparison to the renormalization model. This indicates that applying the multimode model is important for obtaining accurate results.

We also propose that this model exhibits higher numerical accuracy with a given number of computational resources compared to solving the auxiliary-mode model in the flux basis. This claim is supported by two factors. First, solving the auxiliary-mode Hamiltonian becomes in general very challenging for increasing $M > 2$ owing to the exponential increase in the dimension of the computational Hilbert space. Second, as depicted in Fig.~\ref{fig:mode1_anharmonicity}, the results derived from the auxiliary-mode model appear to converge towards those obtained from our multimode model with increasing $M$. However, due to the hefty computational demands imposed by the auxiliary-mode model, we are unable to verify this comprehensively. Note that, in special cases such as $x_\text{J} / l = 0$, we are capable of solving the auxiliary-mode model for $M = 4$ since only auxiliary modes with an even ordinal number exhibit nonzero coupling with the nonlinear mode. In addition, we emphasize that the multimode model requires significantly less computational time than the auxiliary-mode model to achieve a comparable accuracy. More details on the accuracy of the multimode model can be found in Appendix~\ref{app:model_accuracy}.

\section{\label{sec:multimode}Interactions between the lowest modes of the unimon circuit}

\subsection{Energy levels and avoided crossings} \label{subsec:Energy levels and avoided crossings}

Let us leverage the model developed in Sec.~\ref{sec:model} to its full extent by examining the interactions among the three lowest modes of the unimon circuit. This involves solving the multimode Hamiltonian in the unimon basis as given in Eqs.~\eqref{eq:H_nm_matrix}, and subsequently employing our labeling method to map the obtained eigenstates onto the states of the unimon basis. The energies of the labeled eigenstates are denoted as $E_{ijk}$, where $i$, $j$, and $k$ represent the number of excitations in the first, second, and third modes, respectively. In Fig.~\ref{fig:energy_level_spagetti}, we show all obtained energy levels as functions of $x_\text{J}$, $E_\text{J}$, and $Z_\text{c}$. Here, $Z_\text{c} = \sqrt{L_l/C_l}$ represents the characteristic impedance of the transmission line forming the resonator. %Importantly, the sweep of the characteristic impedance is conducted such that the frequency of the lowest auxiliary mode, given by $\Omega_1 = \pi / (2l \sqrt{L_l C_l})$, remains constant. 
Within the energy levels, we highlight the labeled states where each of the initial three modes has at most one excitation. In Fig.~\ref{fig:energy_level_spagetti}(a), the energy levels manifest fairly intricate interactions with each other as a function of $x_\text{J}$. We observe a correlation between the first excited states and the effective charging energies shown in Fig.~\ref{fig:schematic_a_d}(b), which is evident in the alignment of the local extrema. The effective charging energy for the mode depends on the weight coefficient $c_m$ which is related to the coupling between the Josephson junction and mode $m$ [see Eq.~\eqref{eq:Psi_transformed}]. Figs.~\ref{fig:energy_level_spagetti}(b)--(c) illustrate that an increase in either the Josephson energy or the characteristic impedance results in a decrease in energy for the first excited states. This behavior can be attributed to the cancellation of quadratic terms as either $E_\text{J}$ or $Z_\text{c}$ increases. As these quadratic terms decrease, the potential becomes less steep, which in turn narrows the energy gap between the ground state and the excited states. However, the energies of the excited states of the third mode remain largely unchanged, given their weak coupling with the Josephson junction. As discussed in Sec.~\ref{subsec:Labeling of eigenstates}, these interactions often manifest as avoided crossings. An interaction exemplifying this can be observed between energy levels $E_{010}$ and $E_{300}$ at~$x_\text{J}/l = 0.42$, visualized in the inset of Fig.~\ref{fig:energy_level_spagetti}(a).

In qubit operations, avoided crossings are undesirable. If qubit energy levels become hybridized, there is a risk of unintended leakage into other participating states. This can adversely affect the coherence time of the qubit. Consequently, it is preferable to operate in parameter regimes where hybridization of the qubit states is minimized. The prevalence of avoided crossings increases with energy, rendering the high-frequency normal modes ($m \geq 2$) challenging for qubit operation. Nonetheless, in most parameter regimes, the qubit of the lowest mode ($m = 1$) remains unaffected by these avoided crossings up to its second excited state.

\subsection{Anharmonicities and cross-Kerr interactions} \label{subsec:Anharmonicities and cross-Kerr}

\begin{figure}[t]
\includegraphics[scale=0.485]{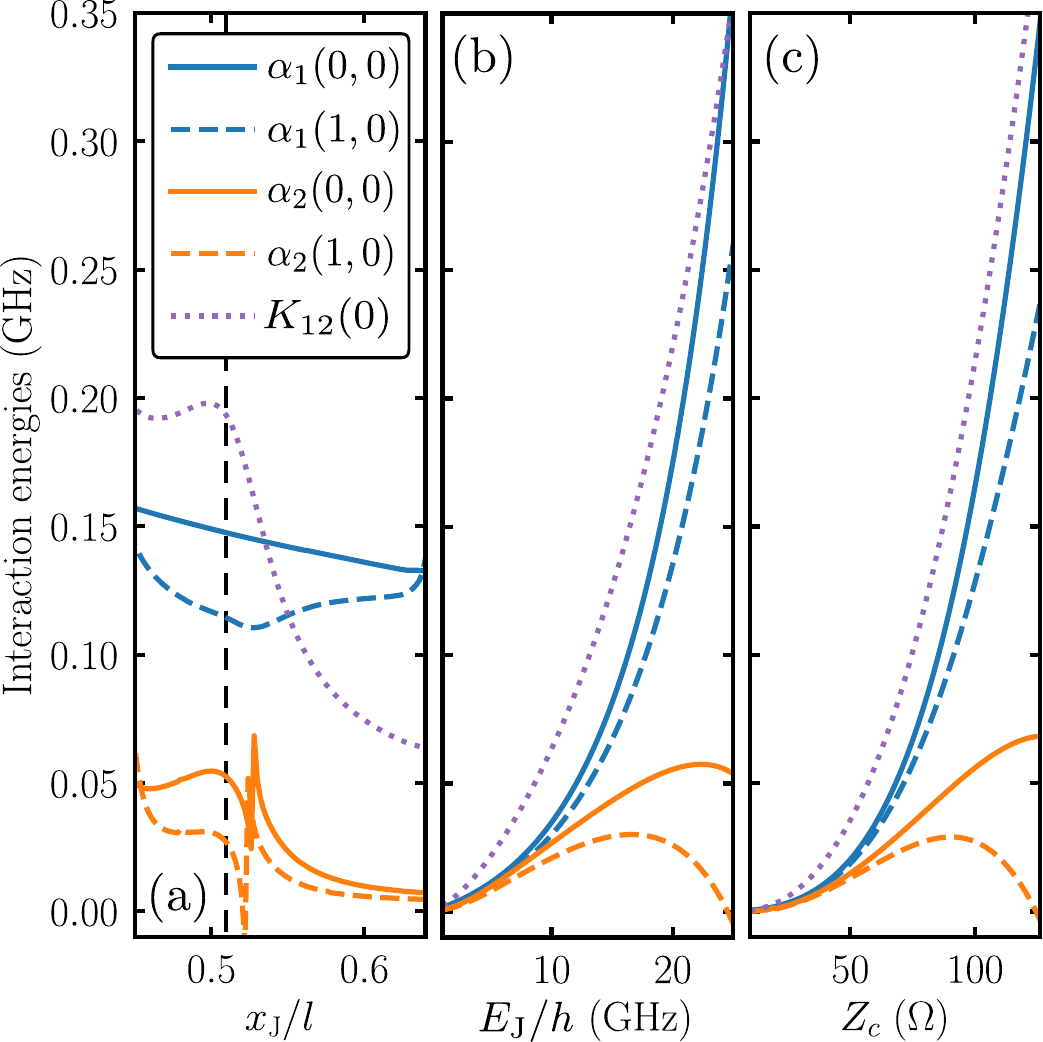}
\caption{\label{fig:interaction_energies_two_mode} Anharmonicities and cross-Kerr interaction energies for the two lowest modes of the unimon circuit as functions of (a)~the location of Josephson junction $x_\text{J}$, (b)~Josephson energy $E_\text{J}$, and (c)~characteristic impedance $Z_\text{c}$ of the CPW. The dashed vertical line in panel (a) indicates the value of $x_\text{J}$, which is subsequently used for the graphs in panels (b) and (c). The notation $\alpha_m (i,j)$ corresponds to the anharmonicity of the $m$:th mode ($m = 1,2,3$), with the other two modes having $i$ and $j$ excitations. The left argument ($i$) pertains to the lower mode of the two arguments. The cross-Kerr interaction is denoted by $K_{nm}(i)$, where $n$ and $m$ ($n,m = 1,2,3$) represent the modes between which the interaction is computed. The argument ($i$) equals to the number of excitations in the mode not primarily involved in the interaction. The simulation utilize the parameter values of Table~\ref{tab:parameter_sets}.}
\end{figure}

\begin{figure*}
\includegraphics[scale=0.55]{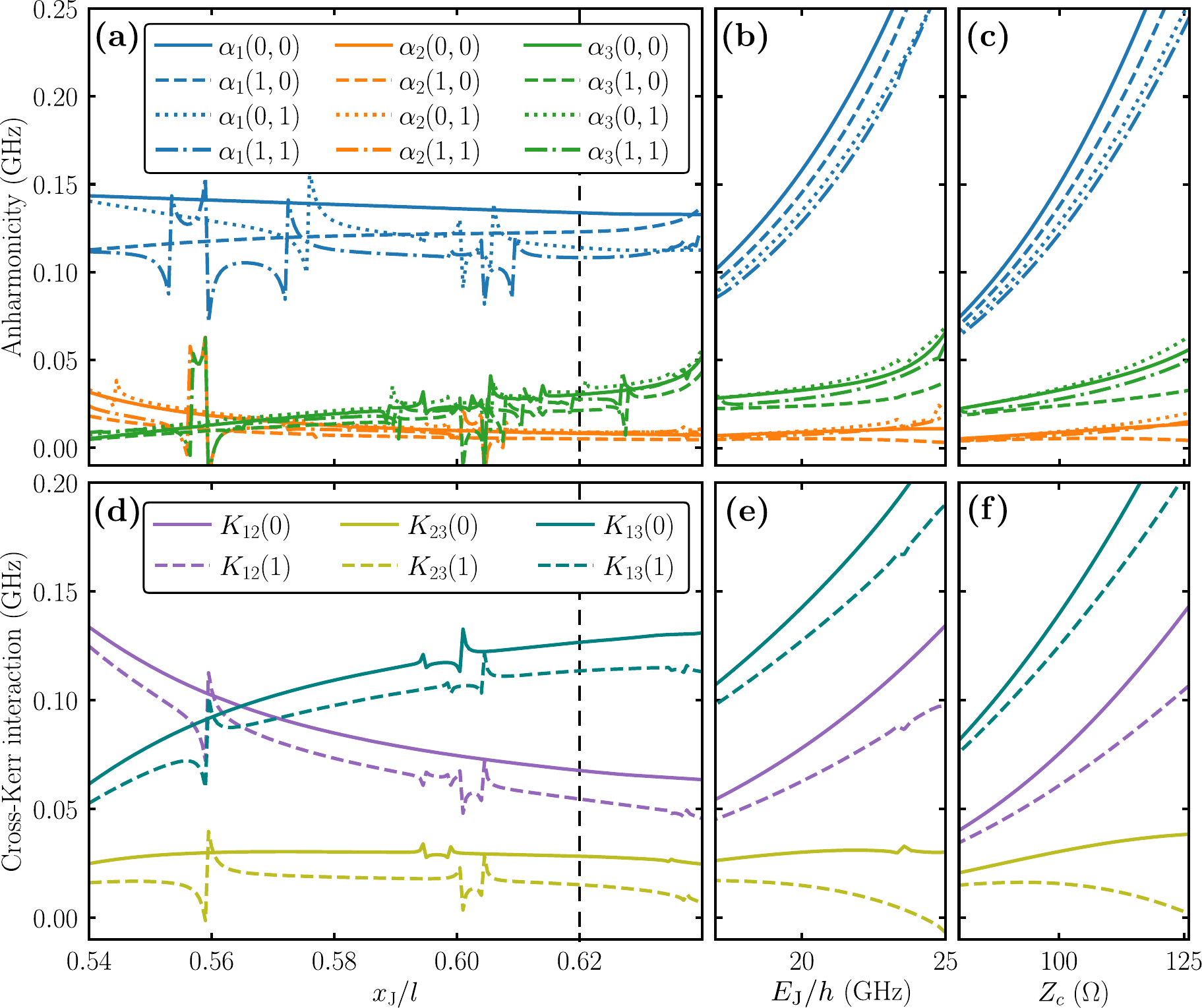}
\caption{\label{fig:interaction_energies_three_mode} (a)--(c) Anharmonicities and (d)--(f) cross-Kerr interaction energies for the three lowest modes of the unimon circuit as functions of (a)~and (d)~the Josephson junction location $x_\text{J}$, (b)~and (e)~Josephson energy $E_\text{J}$, and (c)~and (f)~characteristic impedance of the CPW $Z_\text{c}$. The notation $\alpha_m (i,j)$ and $K_{nm}(i)$ is defined in Eqs.~\eqref{eq:alpha_m} and~\eqref{eq:cross_kerr_mn}, respectively, and in Fig.~\ref{fig:interaction_energies_two_mode}. For the the parameter values that are not swept, we use Table~\ref{tab:parameter_sets}, except for the Josephson junction location set at $x_\text{J} / l = 0.62$, as marked by the vertical dashed line in panels (a) and (d).}
\end{figure*}

We employ the labeled energy levels as a basis to compute key properties, such as the anharmonicity and cross-Kerr interaction. For the lowest mode $m=1$, the anharmonicity is defined as
\begin{equation} \label{eq:alpha_m}
    \alpha_1(i,k) = \frac{(E_{2ik}-E_{1ik})-(E_{1ik}-E_{0ik})}{h},  %= \frac{E_{2ik} - 2 E_{1ik} + E_{0ik}}{h}
\end{equation}
and the cross-Kerr interaction between the modes~1 and~2 is defined as
\begin{equation} \label{eq:cross_kerr_mn}
    K_{12}(i) = \frac{\bigl(E_{11i} - E_{01i}\bigr) - \bigl(E_{10i} - E_{00i}\bigr)}{h}.
\end{equation}
In general, the subscripts in $\alpha_m$ and $K_{mn}$ define which modes are considered to change their excitation number in analogy with the above equations. Although it is customary to consider the anharmonicity and cross-Kerr interaction to involve only one and two modes, respectively, higher-order terms become significant when excitations of the other modes are allowed. In this context, the anharmonicity of a mode depends on the occupation number of the other two modes, while the cross-Kerr interaction between two given modes is influenced by the state of the third mode. We consider these effects in more detail in Sec.~\ref{subsec:Analysis of Kerr-type terms}.

By employing Eqs.~\eqref{eq:alpha_m} and~\eqref{eq:cross_kerr_mn}, we compute the anharmonicity and cross-Kerr interaction for scenarios where only the two lowest modes are allowed a single excitation. Consequently, each mode exhibits two state-dependent anharmonicities and the cross-Kerr interaction between these two modes can be characterized by a single value. The corresponding results for the two-mode scenario are displayed in Fig.~\ref{fig:interaction_energies_two_mode}. Our observations indicate that within the two-mode framework, it is feasible to identify a set of parameters that distribute the nonlinearity relatively evenly across both modes, for example, those given in Table~\ref{tab:parameter_sets}. Under this parameter configuration, if the other mode is in the vacuum state, the anharmonicity for each mode remains above a 50-MHz threshold. Furthermore, the cross-Kerr interaction energy lies around 200 MHz, peaking approximately at the same value of $x_\text{J}$ as $\alpha_2(0,0)$. However, it is evident that the anharmonicity of the lowest mode consistently surpasses that of the second mode. Transferring an excitation to a mode appears to have a diminishing effect on the anharmonicity of the other mode. The relative difference, between the anharmonicities with and without excitations present in the other mode, continues to increase with either increasing Josephson energy or characteristic impedance, even though the overarching trend is an increase in anharmonicity. Intriguingly, the value of $\alpha_1(1,0)$ even begins to decrease once $E_\text{J}$ or $Z_\text{c}$ exceed a certain threshold. More details on this effect is revealed in Sec.~\ref{subsec:Analysis of Kerr-type terms}.

Our numerical calculations also extend to anharmonicities and cross-Kerr interactions for the three-mode scenario, the results of which are illustrated in Fig.~\ref{fig:interaction_energies_three_mode}. In our efforts to spread the nonlinearity across all three modes, we adjusted the location of the Josephson junction to $x_\text{J} = 0.62$ during the sweeps with respect to $E_\text{J}$ and~$Z_\text{c}$. In the case of three modes, the number of anharmonicity parameters per mode increases to four, owing different combinations of excitations in the other two modes. Similarly, the number of cross-Kerr interaction parameters for any pair of modes doubles, reflecting the influence exerted by the state of the remaining mode. The results highlight the challenge in discovering parameters that ensures all anharmonicities exceed a 10-MHz threshold. As discussed in the context of the two-mode case, augmenting $E_\text{J}$ or $Z_\text{c}$ is not a feasible solution as it would result in a decrease in the lowest anharmonicity and cross-Kerr interaction. This outcome is visible for the anharmonicity $\alpha_2(1,0)$ as shown in Fig.~\ref{fig:interaction_energies_three_mode}(b)--(c), and for the cross-Kerr interaction $K_{23}(1)$ as depicted in Fig.~\ref{fig:interaction_energies_three_mode}(e)--(f).

\subsection{Analysis of Kerr-type terms} \label{subsec:Analysis of Kerr-type terms}

Although our model lends itself well to numerical simulations, it does not facilitate analytical derivations. This limitation stems from the fact that the eigenstates of the single-mode unimon Hamiltonian seem not analytically solvable in general. Therefore, to analytically investigate the anharmonicities and cross-Kerr interactions, we make use of the harmonic-oscillator basis~\cite{bourassa_josephson-junction-embedded_2012}. Here, $E_\text{J}^* = E_\text{J} \prod_{m = 1}^{M+1} \ee^{-\lambda_m^2/2}$ represents the renormalized Josephson energy in the harmonic-oscillator basis~\cite{leib_networks_2012,leger_observation_2019}, where $\lambda_m = 2\sqrt{ E_{C,m} / (\hbar \omega_m)}$ signifies the zero-point fluctuations of the $m$:th mode. We refer to the different modes with the indices $n$, $m$, and $k$. %We denote the indices of the different modes as $n$, $m$, and $k$.
Under these definitions, the analytical expressions for the anharmonicity and cross-Kerr interaction are derived as
\begin{align}
\alpha_m(N_n,N_k) = K_{mm} \Bigg( &1 - \sqrt{\frac{2 K_{nn}}{E_\text{J}^*/h}} N_n - \sqrt{\frac{2 K_{kk}}{E_\text{J}^*/h}} N_k \notag \\
&+ \frac{K_{nk}}{E_\text{J}^*/h}  N_n N_k \Bigg), \label{eq:alpha_analytical} \\
K_{nm}(N_k) = K_{nm} \Bigg( & 1 - \sqrt{\frac{2 K_{kk}}{E_\text{J}^*/h}}  N_k \Bigg), \label{eq:K_analytical}
\end{align}
where $N_m$ denotes the occupation number of the $m$:th mode and
%\end{equation}
%Given that other modes are in the vacuum states, 
the self-Kerr and cross-Kerr interaction parameters are given by
\begin{align}
K_{mm} &= \frac{E_\text{J}^* }{2h} \lambda_m^4 = \frac{8 E_\text{J}^* }{h} \left( \frac{E_{C,m}}{\hbar \omega_m} \right)^2, \label{eq:K_mm_analytical} \\
K_{mn} &=  2\sqrt{K_{mm} K_{nn}}, \quad \forall \ m \neq n.\label{eq:K_mn_analytical}
\end{align}
For more details on the derivation of these expressions, refer to Appendix~\ref{app:Analytical calculations}.

\subsubsection{Modes in vacuum state}

We begin our analysis with Eqs.~\eqref{eq:K_mm_analytical} and~\eqref{eq:K_mn_analytical} that represent the case without additional excitations in the system. To gain insight into the behavior of the self-Kerr interaction with respect to $x_\text{J}$, we represent $K_{mm}$ in a more suggestive form, $K_{mm} \propto (E_{C,m} / \omega_m)^2$, where both $E_{C,m}$ and $\omega_m$ are dependent on $x_\text{J}$ as depicted in Fig.~\ref{fig:schematic_a_d}(b)--(c).

Out of these two quantities, the effective charging energy $E_{C,m}$ is notably more sensitive to variations in $x_\text{J}$. As a result, $E_{C,m}$ largely determines the dependence of the anharmonicity of mode $m$ on $x_\text{J}$. Interestingly, the dependence of $E_{C,m}$ on $x_\text{J}$ follows the magnitude of the voltage discontinuity of mode $m$ across the junction, suggesting that $E_{C,m}$ serves as an indicator of the coupling strength between mode $m$ and the Josephson junction.

On the other hand, the transition frequency, $\omega_m/(2 \pi)$, significantly influences the level of the anharmonicity for each mode. A higher transition frequency corresponds to reduced anharmonicity, as depicted in Fig.~\ref{fig:schematic_a_d}(d). Note that we do not consider the dependence of $E_\text{J}^*$ on $x_\text{J}$ in detail because it is relatively weak, as illustrated in Fig.~\ref{fig:mode1_anharmonicity}(b).

The analytical form of the cross-Kerr interaction $K_{nm}$ reveals a simple, yet important relationship with the self-Kerr interactions $K_{mm}$ and $K_{nn}$. Specifically, the cross-Kerr interaction $K_{mn}$ is twice the geometric mean of the self-Kerr interactions $K_{mm}$ and $K_{nn}$. This indicates that the most efficient way to increase the cross-Kerr interaction between the modes, is to augment the self-Kerr interaction of both modes by an equal amount.
%if the nonlinearity is in some sense uniformly distributed between the two modes.
Such conclusions regarding the behavior of the cross-Kerr interaction also applies for the parameter sweeps respect to $E_\text{J}$ and $Z_\text{c}$.

We continue by examining the parameter sweep with respect to the Josephson energy $E_\text{J}$. In this case, the self-Kerr term of mode $m$ can be expressed as $K_{mm} \propto \tilde{E}_\text{J} (E_{C,m}/\omega_m)^2$, highlighting the dependency of the self-Kerr on $E_\text{J}$. From the Fig.~\ref{fig:energy_level_spagetti}(b), we observe that the transition energies from the vacuum state $E_{000}$ to the first excited states $E_{100}$, $E_{010}$, and $E_{001}$ decrease with increasing $E_\text{J}$. Consequently, this leads to an increasing effect on $K_{mm}$, a trend also evident in the numerical results. 

In addition, the renormalized Josephson energy has an enhancing effect on $K_{mm}$, even if the renormalization coefficient may decrease with increasing $E_\text{J}$. As a result, the self-Kerr interaction strength is anticipated to increase more rapidly with respect to $E_\text{J}$ for the lowest mode than for the other modes, a pattern consistent with our numerical calculations.
Another approach to reach this conclusion is by examining the quadratic part of the single-mode Hamiltonian in Eq.~\eqref{eq:Hm}, given by $(E_{L,m} - E_\text{J})\hat{\varphi}_m^2/2$. As $E_\text{J}/E_{L,m} \to 1$, this term approaches zero, while the magnitude of the nonlinear terms grows~\cite{hyyppa_unimon_2022}.

For the sweep with respect to $Z_\text{c}$, we expect similar behavior of the self-Kerr interaction to that for $E_\text{J}$. This is substantiated by the observation that when the operators in the auxiliary-mode Hamiltonian of Eq.~\eqref{eq:H_aux} are rescaled, the outcomes of increasing either $E_\text{J}$ or $Z_\text{c}$ are approximately equal in the linear approximation (Appendix~\ref{app:Sweeping_respect_to}). At the sweet spot, the linear effects are pronounced because the main cause of the increase in self-Kerr interaction is the cancellation of quadratic flux term as $E_\text{J}/E_{L,m} \to 1$. Nevertheless, higher-order terms indicate different behavior between the characteristic impedance $Z_\text{c}$ and the Josephson energy $E_\text{J}$. Such disparities are particularly evident in the case of high-frequency modes, where the effect of the cancellation diminishes. For additional details, refer to Appendix~\ref{app:Sweeping_respect_to}.

\subsubsection{Effect of excitations}

We turn our attention to the impact of excitations on both anharmonicity and cross-Kerr interactions. The origins of these effects can be traced back to nonlinear Hamiltonian terms, which are elaborated upon in Eqs.~\eqref{eq:alpha_analytical} and~\eqref{eq:K_analytical}.

Introducing a single excitation into the $k$:th mode yields two notable outcomes. First, there is a negative correction to the anharmonicity of mode $m$. Second, a corresponding negative correction appears in the cross-Kerr interactions between modes $m$ and $n$. The relative decline for both of these effects is captured mathematically as
\begin{equation} \label{first_correction}
\frac{\alpha_m(1,0) - K_{mm}}{K_{mm}} = \frac{K_{mn}(1) - K_{mn}}{K_{mn}} = - \sqrt{\frac{2 K_{kk}}{E_\text{J}^*/h}},
\end{equation}
which reveals that the decrease caused by the excitation is at its maximum where $K_{kk}/E_\text{J}^*$ attains its peak value. Furthermore, the equation suggests that a mode with a larger self-Kerr interaction induces a larger relative correction. These effects are evident in Fig.~\ref{fig:interaction_energies_two_mode} for the anharmonicity and in Fig.~\ref{fig:interaction_energies_three_mode}(d)--(f) for the cross-Kerr interaction.

Finally, we examine the anharmonicity when another excitation is introduced to the previously unoccupied mode $n$. The difference between this and the preceding case can be expressed as
\begin{equation} \label{second_correction}
\frac{\alpha_m(1,1) - \alpha_m(1,0)}{K_{mm}} = \lambda_n^2 \left( \lambda_k^2 - 1 \right).
\end{equation}
The magnitude of the zero-point fluctuations in mode $k$, which is the mode excited for Eq.~\eqref{first_correction}, determines the sign of the correction. Although the strength of the zero-point fluctuations in mode $n$ does not influence the sign, it acts as a scaling coefficient, influencing the magnitude of the correction. In light of this analysis, the observed trends in Fig.~\ref{fig:interaction_energies_three_mode} align well with our expectations. For a majority of the parameter configurations, the lowest mode shows that $\alpha_1(1,1)$ is the smallest anharmonicity. For the other two modes, we consistently find that $\alpha_m(1,1) > \alpha_m(1,0)$. From a qualitative perspective this makes sense since the lowest mode has the largest zero-point fluctuation. However, quantitatively this behavior is not explained by Eq.~\eqref{second_correction}, since positive correction requires $\lambda_1^2 > 1$ which is not satisfied in this case. %This behavior is in harmony with the predictions from Eq.~\eqref{second_correction}. %We can understand this by considering the fact that excitation in the lowest mode 

It is important, however, to recognize the limits of this analytical methodology. Although instrumental in elucidating the qualitative behavior of Kerr-type interactions, the approach does not encompass transverse-type or other rotating-wave interactions. These interactions may be significant, especially in systems with notable nonlinearity.

\section{\label{sec:conclusions}Conclusions}

In this work, we explored the effects arising from the multimode nature of the single-junction unimon circuit. Whereas our primary focus was on the impact of the high-lying modes on the lowest mode, we also investigated the influence of the nonlinearity on other modes for different locations of the Josephson junction. To facilitate our study, we developed a theory of multimode unimon circuits. Leveraging this framework, we determined the energy spectrum for several low-lying modes using numerical diagonalization in the low-energy subspace.

Our model markedly differs from the one presented in Ref.~\cite{ferguson_symmetries_2013} that describes the multimode physics in Josephson-junction array fluxonium circuits. While the effectiveness of their model relies on the decoupling of the qubit mode from all other modes, the distinctiveness of our model arises from the large energy separation between the qubit subspace and the high-frequency modes, which is a consequence of the CPW structure. Furthermore, the methods of numerical diagonalization for fluxonium qubits presented in Refs.~\cite{hazard_nanowire_2019, smith_quantization_2016} differ from our approach, particularly with respect to the chosen basis. In Ref.~\cite{hazard_nanowire_2019}, diagonalization is executed in the normal-mode flux basis, which enforces a mode cutoff that includes only two modes. On the other hand, the Hamiltonian in Ref.~\cite{smith_quantization_2016} is expressed in the harmonic-oscillator basis, leading to a less efficient convergence of the low-energy eigenstates when compared to our single-mode unimon basis. By adjusting the choice of basis to the problem at hand, we expect that the presented model is also applicable to a class of multimode qubits known as noise-protected qubits~\cite{gyenis_moving_2021,gyenis_experimental_2021,kalashnikov_bifluxon_2020,smith_superconducting_2020,kitaev_protected_2006,weiss_spectrum_2019}.

Our findings reveal that multimode effects introduce significant corrections to both the transition frequency and anharmonicity of the unimon qubit. Utilizing our comprehensive model, the decrease in anharmonicity can be as much as 30\% when the junction is centrally located, with more pronounced effects observed at other positions. We also showed that by solely considering the vacuum states, the multimode effects can be condensed into a single coefficient, leading to a renormalization of the Josephson energy. The decrease of the anharmonicity owing to the renormalization can be compensated by the choice of an increased unnormalized Josephson energy. Compared to the single-mode approach, this method seamlessly incorporates certain multimode effects without greatly increasing the required computational resources. 

We found that the second and third modes present a diverse nonlinearity profile, but their peaks do not coincide. This misalignment poses a challenge in identifying an operational regime where all three of the lowest modes exhibit substantial nonlinearity. Yet, for the first and second modes, we identified a position with notable nonlinearity in both.

Our work significantly expands upon the prior theoretical descriptions of the unimon~\cite{hyyppa_unimon_2022} by introducing a theoretical framework to systematically investigate the multimode physics of the unimon circuit and its potential applications. An interesting application for unimon circuits is to encode multiple qubits into a single device in the spirit of multimon qubits~\cite{roy_multimode_2018, roy_programmable_2020, roy_implementation_2017}. However, although we identified parameter configurations where two modes simultaneously exhibit significant nonlinearity, we did not find promising parameters for particularly high-fidelity qubits. In our future research, we aim to apply our theoretical framework for more complex unimon circuits. For example, by implementing a ring-like geometry and adding more junctions, we expect to witness behavior similar to transmon-based multimons, characterized by a relatively uniform distribution of nonlinearity among the modes with high intrinsic anharmonicity.

\begin{acknowledgments}
The authors acknowledge funding from the Academy of Finland project no. 316619 and the Academy of Finland Centre of Excellence program (project nos. 352925 and 336810), the European Research Council under Advanced Grant no. 101053801 (ConceptQ), FICORE project \emph{Speedup of Quantum Computations by Many-Qubit Logic} under MEC Global program pilot initiatives, and the Vilho, Yrjö and Kalle Väisälä Foundation of the Finnish Academy of Science and Letters.
\end{acknowledgments}

\appendix

\section{Derivation of auxiliary-mode Hamiltonian} \label{app:Derivation}

Here, we derive the auxiliary-mode Hamiltonian that was used as a starting point for the theoretical considerations in Sec.~\ref{sec:model} of the main text.

\subsection{Equations of motion of the distributed-element circuit} 

We initiate our investigation by formulating the total Lagrangian for the continuous distributed-element circuit, presented as
\begin{equation}
    L_\text{tot} = \int_{-l}^{l} \mathcal{L} \left( \psi, \frac{\partial \psi}{\partial x}, \frac{\partial \psi}{\partial t};x,t \right) \dd x + L \left( \Psi, \dot \Psi; t \right),
\end{equation}
where we have defined
\begin{equation} \label{eq:flux_1_2}
\psi(x,t)=
    \begin{cases}
        \psi_1(x,t), & -l < x < x_\text{J}\\
        \psi_2(x,t), & x_\text{J} < x < l
    \end{cases},
\end{equation}
which describes the flux of the grounded CPW resonator and $\Psi(t) = \psi_2(x_\text{J},t) - \psi_1(x_\text{J},t)$ denotes the flux difference across the Josephson junction. The Lagrangian density $\mathcal{L}$ of the CPW and the Lagrangian $L$ describing the flux across the junction are defined as 
\begin{align}
\mathcal{L} = & \frac{C_l}{2} \bigl[\dot{\psi}(x,t) \bigr]^2 - \frac{1}{2 L_l} \bigl[\partial_x \psi(x,t) \bigr]^2 \notag \\
L = & \frac{C_\text{J}}{2} \bigl[\dot{\Psi}(t) \bigr]^2 + E_\text{J} \cos{\biggl\{ \frac{2 \pi}{\Phi_0} \bigl[ \Psi(t) - \Phi_{\text{diff}}  \bigr] \biggr\}}, \label{eq:unimon_lagrangian}
\end{align}
where $\Phi_{\text{diff}}$ is the total flux difference across the loops and the system obeys the boundary condition $\psi(-l,t) = \psi(l,t) = 0$.

To deduce the equations of motion, we employ the Euler-Lagrange equations, defined as
\begin{align}
&\frac{\partial \mathcal{L}}{\partial \psi_i} - \frac{\partial}{\partial t} \frac{\partial \mathcal{L}}{\partial \bigl( \partial \psi_i / \partial t \bigr)} - \frac{\partial}{\partial x} \frac{\partial \mathcal{L}}{\partial \bigl( \partial \psi_i / \partial x \bigr)} = 0, \label{eq:euler_lagrange_field} \\
&\frac{\partial L}{\partial \psi_{\text{J},i}} - \frac{\partial L}{\partial \dot \psi_{\text{J},i}} - (-1)^i \frac{\partial \mathcal{L}}{\partial \bigl( \partial \psi_i / \partial x \bigr)}\bigg|_{x = x_\text{J}} = 0, \label{eq:euler_lagrange_boundary}
\end{align}
where $i \in \left\{ 1,2 \right\}$ and $\psi_{\text{J},i} \equiv \psi_i(x_\text{J},t)$. To utilize the Euler-Lagrange framework, we derive the subsequent equations of motion
\begin{align}
&\partial_x^2 \psi_i(x,t) = \frac{1}{v} \partial_t^2 \psi_i(x,t), \quad i \in \{1,2\}, \label{eq:EoM_WE} \\
&C_\text{J} \ddot{\Psi}(t) + I_\text{c} \sin{\bigg\{ \frac{2 \pi}{\Phi_0} \big[ \Psi(t) - \Phi_{\text{diff}}  \big] \bigg\}} \notag \\
&= \frac{1}{L_l} \partial_x \psi_i(x,t)\big|_{x = x_\text{J}}, \quad i \in \{1,2\}, \label{eq:EoM_xJ}
\end{align}
where the phase velocity is denoted as $v = 1/\sqrt{L_l C_l}$.

\subsection{Variable elimination in frequency domain} 

Our interest is to eliminate the variable $\psi_i$ and express the equation of motion solely with the variable $\Psi$. In order to achieve this, we move to a frequency domain by using Fourier transformation.

Starting with the wave equation in Eq.~\eqref{eq:EoM_WE}, we can express the frequency dependence of the spatial modes $\tilde \psi$ of the CPW resonator as
\begin{align}
\tilde{\psi}_i(x,\omega) = A_i^{(\text{s})}(\omega) \sin(k_{\omega}x) + A_i^{(\text{c})}(\omega) \cos(k_{\omega}x), \label{eq:EoM_WE_freq}
\end{align}
where the wave number is denoted as $k_\omega = \omega / v$.

We can determine the coefficients $A_i^{(\text{s})}$ and $A_i^{(\text{c})}$ by taking the Fourier transformation of Eq.~\eqref{eq:EoM_xJ}, which results in
\begin{equation} \label{eq:EoM_xJ_freq}
\frac{1}{L_l} \partial_x \tilde{\psi}_2(x_\text{J},\omega) = \frac{1}{L_l} \partial_x \tilde{\psi}_1(x_\text{J},\omega).
\end{equation}
This condition ensures that the current across the Josephson junction is continuous. In addition, incorporating the boundary conditions of the grounded CPW and the flux difference across the junction, we can extract the frequency-dependent coefficients as
\begin{align}
A^{(\text{s})}_{1} (\omega) &= -\frac{1}{2} \tilde{\Psi}(\omega) \frac{\cos[k_{\omega}(x_\text{J} - l)]}{\sin(k_{\omega} l)}, \label{eq:A^(s)_1} \\
A^{(\text{c})}_{1} (\omega) &= -\frac{1}{2} \tilde{\Psi}(\omega) \frac{\cos[k_{\omega}(x_\text{J} - l)]}{\cos(k_{\omega} l)}, \label{eq:A^(c)_1} \\
A^{(\text{s})}_{2} (\omega) &= -\frac{1}{2} \tilde{\Psi}(\omega) \frac{\cos[k_{\omega}(x_\text{J} + l)]}{\sin(k_{\omega} l)}, \label{eq:A^(s)_2} \\
A^{(\text{c})}_{2} (\omega) &= \frac{1}{2} \tilde{\Psi}(\omega) \frac{\cos[k_{\omega}(x_\text{J} + l)]}{\cos(k_{\omega} l)}. \label{eq:A^(c)_2}
\end{align}

By inserting the derived coefficients into Eqs.~\eqref{eq:EoM_WE_freq} and~\eqref{eq:EoM_xJ_freq}, we obtain
\begin{equation} \label{eq:EoM_xJ_kernel}
\frac{1}{L_l} \partial_x \tilde{\psi}_i(x_\text{J},\omega) = -\tilde{K}(\omega) \tilde{\Psi}(\omega),
\end{equation}
where the kernel function is defined as
\begin{equation} \label{eq:K_kernel_freq}
\tilde{K}(\omega) = \frac{k_{\omega} \cos \bigl[k_{\omega}(x_\text{J} + l)\bigr] \cos \bigl[k_{\omega}(x_\text{J} - l) \bigr]}{L_l \sin \bigl(2 k_{\omega}l \bigr)}.
\end{equation}
This kernel $\tilde K$ encapsulates the intricate relationship between the Josephson junction and the CPW resonator, enabling us to focus solely on the $\Psi$ variable.

Switching back to the time domain with the help of the inverse Fourier transformation, gives rise to the integro-differential equation
\begin{align}
&C_\text{J} \ddot{\Psi}(t) + I_\text{c} \sin{\bigg\{ \frac{2 \pi}{\Phi_0} \big[ \Psi(t) - \Phi_{\text{diff}}  \big] \bigg\}} \notag \\
&+ \int_{-\infty}^{\infty} K(t - \tau) \Psi(\tau) \dd \tau = 0. \label{eq:integro_differential}
\end{align}
Although this equation provides a stepping stone into the classical dynamics of the system, our end goal is to formulate the quantized Hamiltonian. For this purpose, we continue in the frequency domain, aiming to replace the kernel function with resonator eigenmodes.

\subsection{Pole expansion of the kernel function}

To carry out a pole decomposition of the kernel $\tilde K$, we treat $\omega$ as a complex variable. This lets us use a Mittag-Leffler variant to express the kernel function. The decomposition is expressed as
\begin{align}
&\tilde{K}(\omega) = \sum_{n=0}^{\infty} \frac{\dd^n \tilde{K}(\omega)}{\dd \omega^n}\bigg|_{\omega = 0}\frac{\omega^n}{n!} \label{eq:Kernel_pole_expansion} \\
&+ \sum_{m = 1}^M \bigg\{  \frac{2 r_m \Omega_m}{\omega^2 - \Omega_m^2} + \frac{r_m}{\Omega_m}\sum_{n=0}^{\infty}  [1 + (-1)^n ] \Big( \frac{\omega}{\Omega_m} \Big)^n \bigg\}, \notag
\end{align}
where $M$ is a positive integer that denotes the number modes and the poles $\Omega_m = \pi v m / (2l)$ correspond to the harmonic modes of the resonator as can be observed from Eq.~\eqref{eq:K_kernel_freq}. The residues for each pole are denoted as $r_m = \text{res}[\tilde{K}(\omega), \Omega_m]$. The first summation in Eq.~\eqref{eq:K_kernel_freq} represents the Maclaurin series, providing an estimate of $\tilde K(\omega)$ at low frequencies. The second summation term is introduced by adding zeros in a form of geometric series as
\begin{equation} \label{eq:geometric_series}
0 = \frac{r_m}{\omega - \Omega_m} + \frac{r_m}{\Omega_m} \sum_{n = 0}^{\infty} \Big( \frac{\omega}{\Omega_m} \Big)^n,
\end{equation}
which is valid under the condition $\abs{\omega / \Omega_m} < 1$. The residues of each pole are given by
\begin{equation} \label{eq:residue}
r_{m} = \frac{\Omega_m \cos^2 \Big( \frac{\Omega_m x_\text{J}}{v} + \frac{\pi}{2}m \Big)}{2 l L_l},
\end{equation}
which satisfies $r_{m} = -r_{-m}$.

\subsection{Introducing auxiliary modes}

Using the pole decomposition form of the kernel function, the convolution term in Eq.~\eqref{eq:integro_differential} can be expressed as
\begin{align} \label{eq:convolution}
&\int_{-\infty}^{\infty} K(t - \tau) \Psi(\tau) \dd \tau=\mathcal{F}^{-1} \Biggl[ \sum_{m = 1}^M \frac{2 r_m \Omega_m}{\omega^2 - \Omega_m^2}\tilde{\Psi} (\omega)\Biggr](t) \notag \\
& \sum_{\substack{n = 0 \\ \text{even $n$}}}^{\infty} \Bigg\{ \frac{1}{n!} \frac{\dd^n \tilde{K}(\omega)}{\dd \omega^n}\bigg|_{\omega = 0} + \sum_{m = 1}^M \frac{2 r_m}{\Omega_{m}^{n+1}} \Bigg\} (-i)^{n} \partial_t^n \Psi(t) \notag \\
&\approx \Bigg[ \tilde{K}(0) + \sum_{m = 1}^M \frac{ 2 r_m}{\Omega_{m}} \Bigg] \Psi(t) - \Bigg[ \frac{\tilde{K}''(0)}{2} + \sum_{m = 1}^M \frac{2 r_m}{\Omega_{m}^{3}} \Bigg] \ddot{\Psi}(t) \notag \\
&+ \mathcal{F}^{-1} \Bigg[ \sum_{m = 1}^M \frac{2 r_m \Omega_m}{\omega^2 - \Omega_m^2}\tilde{\Psi}(\omega) \Bigg](t),
\end{align}
where $\mathcal{F}^{-1}$ denotes the inverse Fourier transform. We have also used the convolution theorem and the fact that contribution from odd time-derivatives of flux is zero. This is evidenced by the fact that, if $n$ is odd, both $\dd^n \tilde{K}(\omega) / (\dd \omega^n)\big|_{\omega = 0}$ and the latter term in the second row of Eq.~\eqref{eq:Kernel_pole_expansion} are zero. 

In addition, since the only lumped element which depends on time-derivatives of flux is capacitance, we may utilize the low-frequency assumption made with the pole expansion to approximate the convolution by neglecting time-derivatives $\partial_t^n \Psi(t)$ when $n > 2$. Interestingly, this low-frequency approximation made in the convolution expression in Eq.~\eqref{eq:convolution} becomes exact in the limit of $M~\to~\infty$. In this limit, Mittag-Leffler theorem~\cite{arfken_chapter_2013} allows us to express the kernel in Eq.~\eqref{eq:Kernel_pole_expansion} as
\begin{equation}
    \tilde{K}(\omega) = \tilde{K}(0) + \sum_{m = 1}^\infty \bigg\{  \frac{2 r_m \Omega_m}{\omega^2 - \Omega_m^2} + \frac{2 r_m}{\Omega_m} \bigg\},
\end{equation}
where all contributions from kernel $\tilde K(\omega)$ to the time-derivatives in the convolution term vanish. Similar technique has been previously utilized in the context of network synthesis of prescribed impedance functions~\cite{zinn_network_1952}.

The approximation of convolution, as given in Eq.~\eqref{eq:convolution}, leads us to a more tractable expression for the full equation of motion in Eq.~\eqref{eq:integro_differential} which takes the form
\begin{align}
&C_\text{eff} \ddot{\Psi}(t) + L_\text{eff}^{-1}\Psi(t) + I_\text{c} \sin{\bigg\{ \frac{2 \pi}{\Phi_0} \big[ \Psi(t) - \Phi_{\text{diff}}  \big] \bigg\}} \notag \\
&= -\mathcal{F}^{-1} \Bigg[ \sum_{m = 1}^M \frac{2 r_m \Omega_m}{\omega^2 - \Omega_m^2}\tilde{\Psi} (\omega)\Bigg](t). \label{eq:EoM_nonlocal}
\end{align}
where we used the definition for $C_\text{eff}$ given in the Eq.~\eqref{eq:C_eff} of the main text and defined
\begin{equation} \label{eq:L_eff}
    L_{\text{eff}} = \left[ \frac{1}{2 l L_l}  + \sum_m^M \frac{\cos^2 \bigl( \frac{\Omega_m x_\text{J}}{v} + \frac{m \pi}{2} \bigr)}{l L_l} \right]^{-1}.
\end{equation}

To handle the temporally nonlocal term, we employ a set of auxiliary modes, denoted as
\begin{equation} \label{eq:auxiliary_modes}
\chi_m(t) =  \mathcal{F}^{-1} \Bigg[\frac{\xi_m \tilde{\Psi}(\omega)}{\omega^2 - \Omega_m^2} \Bigg](t),
\end{equation}
where $\xi_m = \sqrt{2 r_m \Omega_m C_\text{eff}}$. This helps us to express Eq.~\eqref{eq:EoM_nonlocal} as
\begin{align}
&C_\text{eff} \ddot{\Psi}(t) + L_\text{eff}^{-1}\Psi(t) + \sum_{m = 1}^M \xi_m \chi_m(t) \notag \\
&= -I_\text{c} \sin{\bigg\{ \frac{2 \pi}{\Phi_0} \big[ \Psi(t) - \Phi_{\text{diff}}  \big] \bigg\}}. \label{eq:EoM_Psi}
\end{align}
For a complete description of the system dynamics, we derive the equations for each auxiliary mode. By utilizing Eq.~\eqref{eq:auxiliary_modes} and the convolution theorem, we obtain
\begin{equation} \label{eq:EoM_auxiliary}
C_\text{eff} \ddot{\chi}_m(t) + \xi_m \Psi(t) + C_\text{eff} \Omega_m^2 \chi_m(t) = 0.
\end{equation}
By introducing auxiliary modes, we have successfully eliminated the need to compute temporal convolutions.

\subsection{Classical treatment of the dc flux}

In the pursuit of computing the normal modes of the system, it is convenient to redefine the fluxes such that they vanish at the minima of their effective potentials. To this end, we define the shifted flux and the auxiliary variables as follows: $\Psi(t)' = \Psi(t) - \Phi_\text{diff} + \phi_0$ represents the total flux as a sum of its dynamic component $\Psi'(t)$ and a shift $\phi_0$. Similarly, for the auxiliary modes, we define $\chi_m'(t) = \chi_m(t) + x_m$.

Substituting these into our earlier equations, we obtain
\begin{align}
&C_\text{eff} \ddot{\Psi}'(t) + L_\text{eff}^{-1}\Psi'(t) + \sum_{m = 1}^M \xi_m \chi_m'(t) \notag \\
&+ I_\text{c} \sin{\bigg\{ \frac{2 \pi}{\Phi_0} \big[ \Psi'(t) - \phi_0 \big] \bigg\}} + I_\text{c} \sin\biggl( \frac{2 \pi \phi_0}{\Phi_0} \biggr)\label{eq:EoM_Psi'} \\
=&- L_\text{eff}^{-1} \bigl( \Phi_\text{diff} - \phi_0 \bigl) - \sum_{m = 1}^M \xi_m x_m + I_\text{c} \sin\biggl( \frac{2 \pi \phi_0}{\Phi_0} \biggr) \notag
\end{align}
\begin{align}
&\ddot{\chi}_m'(t) + \xi_m \Psi'(t) + C_\text{eff} \Omega_m^2 \chi_m'(t) \notag \\
=& -\xi_m\bigl( \Phi_\text{diff} - \phi_0 \bigl) - C_\text{eff} \Omega_m^2 x_m. \label{eq:EoM_auxiliary'}
\end{align}
It is clear that the right side of these equations captures the time-independent behavior, with the left side containing time-dependent part of the system.

Conveniently, the time-independent parts of Eqs.~\eqref{eq:EoM_Psi'} and~\eqref{eq:EoM_auxiliary'} can be solved independently from the time-dependent parts, giving us a set of new equations 
\begin{align}
& I_\text{c} \sin\biggl( \frac{2 \pi \phi_0}{\Phi_0} \biggr) - \frac{\Phi_\text{diff} - \phi_0 }{L_\text{eff}} = \sum_{m = 1}^M \xi_m x_m \\
& C_\text{eff} \Omega_m^2 x_m = -\xi_m\bigl( \Phi_\text{diff} - \phi_0 \bigl).
\end{align}
From these equations, by eliminating the variables $x_m$, we obtain the relation
\begin{equation} \label{eq:dc_phase}
\frac{ \Phi_\text{diff} - \phi_0 }{2 l L_l} - I_\text{c} \sin\biggl( \frac{2 \pi \phi_0}{\Phi_0} \biggr) = 0.
\end{equation}
This equation links the dc flux across our junction to the external magnetic flux, encapsulating the dc flux behavior of our system in the presence of an external magnetic influence. Note that the Eq.~\eqref{eq:dc_phase} becomes multi-valued if $2 l L_l / L_\text{J} > 1$~\cite{miano_hamiltonian_2023, rymarz_consistent_2023, weiss_variational_2021}, a parameter region we aim to avoid.

\subsection{Finding the Hamiltonian and quantization}

Let us derive the Lagrangian based on the left side of Eqs.~\eqref{eq:EoM_Psi'} and~\eqref{eq:EoM_auxiliary'}. The Lagrangian is defined as
\begin{align}
\mathcal{L}_\text{aux} = &\frac{C_\text{eff} \dot{\Psi}'^2}{2} - \frac{\Psi'^2}{2 L_\text{eff}} + E_\text{J} \cos\left(\frac{2 \pi}{\Phi_0} [ \Psi' - \phi_0 ]\right) \notag \\
&+ \sum_{m = 1}^M C_\text{eff} \left[ \frac{\dot{\chi}_m'^2}{2} - \Omega_m^2 \frac{\chi_m'^2}{2} \right] - \sum_{m = 1}^M \xi_m \chi_m' \Psi' \notag \\
&- \frac{\Psi'}{2 l L_l}(\Phi_{\text{diff}} - \phi_0),
\end{align}
where we omitted the explicit temporal dependencies for brevity.

In order to transition to the Hamiltonian formalism, we define the conjugate momenta associated with $\Psi'$ and $\chi_m'$ as
\begin{align}
&Q' = C_\text{eff} \dot{\Psi}', \
&\Xi_m' = C_\text{eff} \dot{\chi}_m'.
\end{align}
Using a Legendre transformation and subsequent quantization, we obtain
\begin{align} 
\hat{H}_\text{aux} = &\frac{\hat{Q}'^2}{2 C_\text{eff}} + \frac{\hat{\Psi}'^2}{2 L_\text{eff}} + \frac{\hat{\Psi}'}{2 l L_l}(\Phi_{\text{diff}} - \phi_0) \notag \\
&+ \sum_{m = 1}^M \bigg[ \frac{\hat{\Xi}_m'^2}{2 C_\text{eff}} + C_\text{eff}\Omega_m^2 \frac{\hat{\chi}_m'^2}{2} \bigg] + \sum_{m = 1}^M \xi_m \hat{\chi}_m' \hat{\Psi}' \notag \\
& - E_\text{J} \cos{\bigg[ \frac{2 \pi}{\Phi_0} \big( \hat{\Psi}' - \phi_0  \big) \bigg]}, \label{eq:H_aux_v1}
\end{align}
where the associated quantum operators obey the commutation relations $[\hat{\Psi}', \hat{Q}'] = \ii\hbar$ and $[\hat{\chi}_n', \hat{\Xi}_m'] = \ii\hbar \delta_{nm}$ with all other commutators giving zero. This Hamiltonian can be expressed in a more compact form as
\begin{align} 
\hat{H}_\text{aux} = &\frac{\hat{Q}'^2}{2 C_\text{eff}} + \frac{1}{4 l L_l} \Bigl( \hat \Psi' + \Phi_\text{diff} - \phi_0\Bigr)^2 \notag \\
&+ \sum_{m = 1}^M \Bigg[ \frac{\hat{\Xi}_m'^2}{2 C_\text{eff}} + \frac{1}{2}C_\text{eff}\Omega_m^2 \biggl( \hat{\chi}_m' + \frac{\xi_m \hat \Psi'}{C_\text{eff}\Omega_m^2} \biggr)^2 \Bigg] \notag \\
&- E_\text{J} \cos{\bigg[ \frac{2 \pi}{\Phi_0} \big( \hat{\Psi}' - \phi_0  \big) \bigg]},
\end{align}
after the static terms are discarded. Note that $L_\text{eff}$ in the second term of Eq.~\eqref{eq:H_aux_v1} is expressed explicitly by using Eqs.~\eqref{eq:L_eff} and~\eqref{eq:xi_m}. Omission of primes in this Hamiltonian leads to the auxiliary-mode Hamiltonian in Eq.~\eqref{eq:H_aux}.

\section{Linearization of auxiliary-mode Hamiltonian} \label{app:Linearization}

In this appendix, we supplement the derivation of the normal-mode representation of the auxiliary-mode Hamiltonian $\hat H_\text{aux}$ with details that were omitted in the Sec.~\ref{sec:Effective multimode unimon Hamiltonian} of the main text.

The linear and nonlinear parts of the auxiliary-mode Hamiltonian are defined as
\begin{align}
     \hat{H}_{\text{lin}} = &\frac{\hat{Q}^2}{2 C_{\text{eff}}} + \frac{1}{2} \biggl[ \frac{1}{L_\text{eff}} + \frac{\cos(2 \pi \phi_0 / \Phi_0)}{L_\text{J}} \biggr] \hat{\Psi}^2 \notag \\
    & +\sum_{m = 1}^M \biggl[ \frac{\hat{\Xi}_m^2}{2 C_{\text{eff}}} + \frac{1}{2} C_{\text{eff}} \Omega_m^2 \hat{\chi}_m^2 + \xi_m \hat{\chi}_m \hat{\Psi} \biggr], \label{eq:H_lin}\\
     \hat{H}_{\text{nl}} = &- E_{\text{J}} \cos \biggl[ \frac{2 \pi}{\Phi_0} ( \hat{\Psi} - \phi_0 ) \biggr] + I_\text{c} \sin \biggl( \frac{2 \pi \phi_0}{\Phi_0} \biggr) \hat{\Psi}  \notag\\
    &- \frac{1}{2 L_\text{J}} \cos \biggl( \frac{2 \pi \phi_0}{\Phi_0} \biggr) \hat{\Psi}^2, \label{eq:H_nl}
\end{align}
where the last two terms cancel out the first- and second-order contributions from the cosine function, consequently showing that $\hat H_{\rm nl}$ contributes to fully nonlinear dynamics. This becomes particularly clear if the trigonometric identity $\cos (x - y) = \cos x \cos y + \sin x \sin y$ is applied. We also note that the dc magnetic flux offset across the junction $\phi_0$ and the half difference of the external magnetic flux $\Phi_\text{diff}$ are connected through a relation
\begin{equation} \label{eq:dc_flux}
\frac{\Phi_{\text{diff}} - \phi_0}{2 l L_l} - I_\text{c} \sin \biggl( \frac{2 \pi \phi_0}{\Phi_0} \biggr) = 0,
\end{equation}
which is used in Eq.~\eqref{eq:H_lin}. As elaborated upon in Appendix~\ref{app:Derivation}, this relation arises from the time-independent part of the classical Lagrange equation for the system. Specifically, it demonstrates how the dc flux difference across the Josephson junction is influenced by the difference in external fluxes passing through the two loops.

In a matrix form the linear part is expressed as
\begin{equation} \label{eq:H_lin_matrix}
\hat H_\text{lin} =  \frac{1}{2} \mathbf Q^{\dagger} \mathbf C^{-1} \mathbf Q + \frac{1}{2} \mathbf V^{\dagger} \mathbf L^{-1} \mathbf V,
\end{equation}
where the flux vector $\mathbf V$ is defined as $[\mathbf V]_{0} = \hat \Psi$, $[\mathbf V]_{m} = \hat \chi_m, \ m \in \{ 1, 2, ..., M \}$ and correspondingly the charge vector $\mathbf Q$ is defined as $[\mathbf Q]_{0} = \hat Q$, $[\mathbf Q]_{m} = \hat \Xi_m, \ m \in \{ 1, 2, ..., M \}$. Since there is no charge coupling present in the system, the inverse of the capacitance matrix is simply $[\mathbf C^{-1}]_{nn} = 1 / C_\text{eff}, \ n \in \{ 0, 1, 2, ..., M \}$. For the inverse of the inductance matrix, the nonzero elements are defined as $[\mathbf L^{-1}]_{00} = L_\text{eff}^{-1} + \cos(\varphi_0)/L_\text{J}$, $[\mathbf L^{-1}]_{mm} = C_\text{eff} \Omega_m^2$ and $[\mathbf L^{-1}]_{0m} = [\mathbf L^{-1}]_{m0} = \xi_m, \text{for} \ m \in \{ 1, 2, ..., M \}$.

Such matrix can be diagonalized by introducing an unitary matrix $\mathbf U$ that satisfies $\mathbf D = \mathbf U^\text{T} \mathbf L^{-1} \mathbf U$, where the matrix $\mathbf D$ is diagonal. By utilizing this, the Hamiltonian in Eq.~\eqref{eq:H_lin_matrix} can be expressed as 
\begin{align} 
\hat H_\text{lin} = \frac{1}{2} \mathbf q^{\dagger} \mathbf C^{-1} \mathbf q + \frac{1}{2}\mathbf v^{\dagger} \mathbf D \mathbf v,
\end{align}
where $[\mathbf q]_{n} = [\mathbf U^{\dagger} \mathbf Q]_{n} = \hat q_n, \ n \in \{0, 1, 2, ..., M \}$ and $[\mathbf v]_{n} = [\mathbf U^{\dagger} \mathbf V]_{n} = \hat \phi_n, \ n \in \{0, 1, 2, ..., M \}$.

Finally, the linear part of the Hamiltonian can be expressed as
\begin{equation}
H_\text{lin} =  \sum_{m = 0}^M \Bigg[\frac{q_m^2}{2 C_\text{eff}} + \frac{1}{2} C_\text{eff} \omega_m^2\phi_m^2 \Bigg],
\end{equation}
where the normal-mode frequencies are denoted as $\omega_m/(2 \pi)$. Furthermore, the normal-mode decompositions for operators $\hat \Psi$ and $\hat \chi_m$ take the form
\begin{equation}
\hat{\Psi} = \sum_{m = 1}^{M+1} [ \mathbf U ]_{0,m-1} \hat{\phi}_m, \quad \hat{\chi}_m = \sum_{n = 1}^{M+1} [ \mathbf U ]_{m,n-1} \hat{\phi}_n,
\end{equation}
and thus, the coefficients $c_m$ in Eq.~\eqref{eq:Psi_transformed} of the main text are given by $c_m = [ \mathbf U ]_{0,m-1}$.

\section{Division to single-mode and interaction parts} \label{app:division_to_Hm_and_H_int}

%In Eq.~\eqref{eq:Hm_plus_Hint} of the main text, we separate the normal-mode Hamiltonian into two distinct parts: the single-mode and the interaction. However, based on our chosen representation, it might not be immediately clear that the interaction part exclusively comprises interaction terms.

Starting with the assumption that $\varphi_0 = \pi$ and inspecting Eq.~\eqref{eq:cos_infinite_sum} of the main text, we note that only the term containing only cosines is a single-mode term. All other terms involve products of sines, which introduce coupling between the modes. We proceed by expressing the product of cosines as
\begin{align}
&\prod_{i = 1}^N \bigl[1 + \bigl( \cos \hat \varphi_i - 1 \bigr) \bigr] \\
&= 1 + \sum_{k = 1} ^{N} \bigl(\cos \hat \varphi_i -1 \bigr)+ \sum_{k = 2} ^{N} \sum_{\substack{A\subseteq\{1,...,N \} \\ \abs{A} = k}} \prod_{i \in A} \bigl( \cos \hat \varphi_i - 1 \bigr), \notag
\end{align}
The significance of $\bigl( \cos \hat \varphi_m - 1 \bigr)$ arises from the cancellation of the constant term. Consequently, the last summation term inevitably includes only interaction terms, whereas the second summation term holds the single-mode contribution. By applying the above steps to Eq.~\eqref{eq:H_int}, the single-mode contributions conveniently cancel each other out, and hence only interaction terms remain.
%\section{Derivation of cosine of sums} \label{app:cosine_of_sums}
%\section{Equivalence between two models} \label{app:Equivalence}
%asd

\section{Scaling of the renormalization coefficient} \label{app:scaling}

Here, we study the scaling properties of the renormalization coefficient $\tilde E_\text{J} / E_\text{J}$ in more detail. We put forward a convincing argument that the mode cutoff frequency used in the multimode model of the main text is reasonable.

\begin{figure}[t]
\includegraphics[scale=0.47]{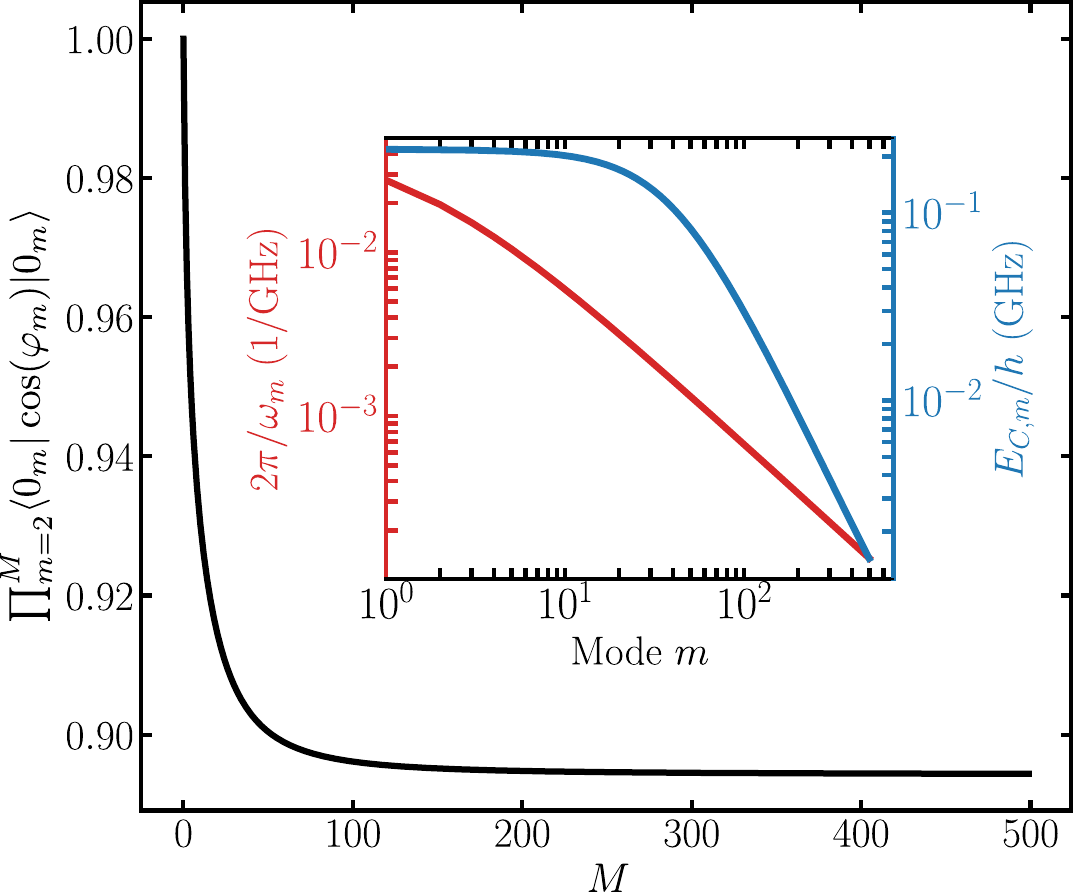}
\caption{\label{fig:renorm_scaling} Renormalization coefficient $\tilde E_\text{J} / E_\text{J}$ ($M_0 = 1$) as a function of the total number of modes $M$ (black curve). The inset shows the inverse of the mode eigenfrequency (red curve) and effective charging energy $E_{C,m} = c_m^2 e^2/(2 C_\text{eff})$ (blue curve) as functions of the mode number $m$. The calculations are carried out for $x_\text{J} / l = 0$ and $\phi_0  = \pi$. Note that only modes that couple to the Josephson junction are included. In this case only every other mode is coupled.}
\end{figure}

In Fig.~\ref{fig:renorm_scaling}, we show the renormalization coefficient for the lowest mode as a function of the total number of modes that are coupled to the Josephson junction. Importantly, the renormalization coefficient appears to converge to a value slightly below $0.90$. This supports our choice of the mode cutoff frequency, based on the superconducting gap, which yielded a renormalization coefficient of approximately $\tilde E_\text{J} / E_\text{J} \approx 0.90$ (see Sec.~\ref{sec:single-mode} in the main text).

To understand this behavior, we consider the renormalization coefficient expressed in the harmonic-oscillator basis as
\begin{equation} \label{eq:renorm_HO_basis}
E_\text{J}^* / E_\text{J} = \prod_{m = 2}^{M} \ee^{-\lambda_m^2/2}, \ \ \ \lambda_m = 2\sqrt{\frac{E_{C,m}}{\hbar \omega_m}}
\end{equation}
where $\lambda_m$ denotes the zero-point fluctuations of the mode $m$. In the exponent, both the inverse of the mode angular eigenfrequency and the effective charging energy are present, which are both depicted in the inset of Fig.~\ref{fig:renorm_scaling} as functions of mode number $m$. Due to the CPW structure, the eigenfrequency scales as $ \sim 1/m$. This scaling alone would lead to $E_\text{J}^* / E_\text{J} \sim 1/ M$, causing the renormalization coefficient to vanish in the limit $M \to \infty$. However, the effective charging energy appears to scale as $\sim 1 / m^{\gamma}$, where $\gamma > 1$. Note that there is a clear change in the scaling behavior around $m \approx 20$, after which the effective charging energy decreases faster than the inverse eigenfrequency, implying that $\gamma > 1$ in the limit $m \to \infty$. In the context of the exponent in Eq.~\eqref{eq:renorm_HO_basis}, this results in a scaling given by $\lambda_m \sim (1/m)^{1 + \gamma}$. Should the assumption $\gamma > 1$ hold true, it implies that the renormalization coefficient converges to a nonzero value, consistent with the behavior observed in Fig.~\ref{fig:renorm_scaling}.

Similar arguments regarding the convergence of the renormalization effects, especially when considering a Josephson junction that is capacitively coupled to the end of a CPW, are discussed in Refs.~\cite{gely_convergence_2017,parra-rodriguez_quantum_2018,malekakhlagh_cutoff-free_2017}. These works demonstrated that the magnitude of the coupling between the qubit mode and the high-frequency modes exhibits a natural cutoff frequency. This cutoff is exclusively dependent on the Josephson capacitance $C_\text{J}$ as $M \to \infty$. We observed similar behavior in the mode-dependent effective charging energies of the unimon circuit. Specifically, in the inset of Fig.~\ref{fig:renorm_scaling}, a natural frequency cutoff is apparent around $M \approx 50$. Although not shown in the figure, this cutoff is strongly influenced by the selected value of $C_\text{J}$.

\section{Accuracy of the multimode model} \label{app:model_accuracy}

\begin{figure}[t]
\includegraphics[scale=0.47]{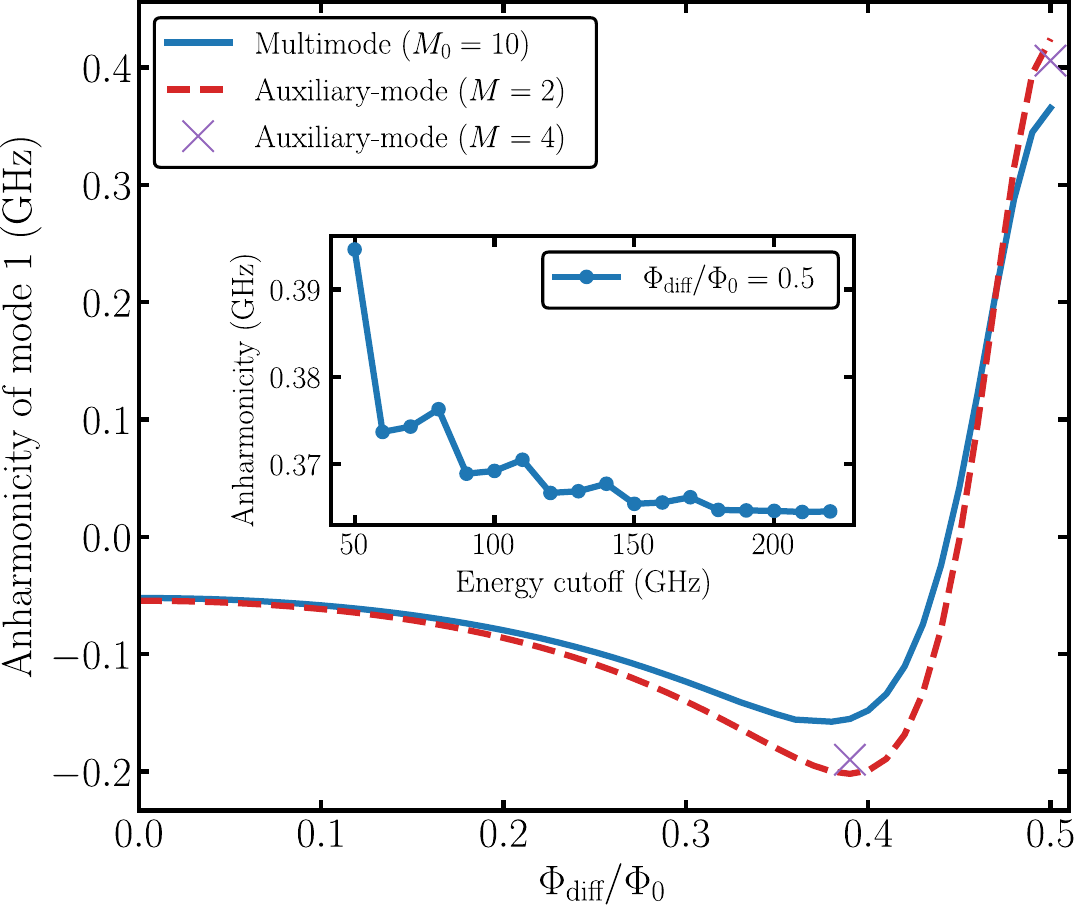}
\caption{\label{fig:multimode_Phi_sweep_convergence} Anharmonicity of the lowest mode as a function of the $\Phi_\text{diff} / \Phi_0$ for the multimode ($M_0 = 10$) and auxiliary-mode ($M = 2$) models. In addition, two data points for the auxiliary-mode model with $M = 4$ are shown at positions where the deviation between the two models is most pronounced ($\Phi_\text{diff} / \Phi_0 = 0.39$ and $0.50$). The inset shows the anharmonicity of lowest mode as a function of the energy cutoff $E_\text{cutoff}$, as determined by the multimode model. The parameters used for these calculations are $x_\text{J} / l = 0$, $L_l = 0.83$ $\upmu$H/m, $C_l = 83.0$ pF/m, and $E_\text{J} / h = 19.0$ GHz.}
\end{figure}

In this appendix, we address the accuracy of the multimode model compared with the auxiliary-mode model which are described by the Hamiltonians given in Eqs.~\eqref{eq:H_nm_final} and~\eqref{eq:H_aux}, respectively. Figure~\ref{fig:multimode_Phi_sweep_convergence} illustrates that significant discrepancies in the lowest-mode anharmonicity between the two models arise primarily at two flux bias points: $\Phi_\text{diff} / \Phi_0 = 0.39$ and $0.50$. We observe that if the number of auxiliary modes increases from $M = 2$ to $M = 4$, the anharmonicity undergoes corrections of approximately $15$ MHz in magnitude. In both instances, these corrections move the value closer to what is obtained with the multimode model. In addition, the inset of Fig.~\ref{fig:multimode_Phi_sweep_convergence} reveals a convergent behavior for the multimode model, exhibiting markedly smaller deviations than the auxiliary-mode model as the energy cutoff $E_\text{cutoff}$ surpasses $100$ GHz. For instance, if $E_\text{cutoff}$ is increased from 150 GHz to 200 GHz, the corrections to the anharmonicity is roughly 1 MHz.

\section{Dependence of the unimon physics on \texorpdfstring{$E_\text{J}$}{EJ} and \texorpdfstring{$Z_\text{c}$}{Zc}} \label{app:Sweeping_respect_to}

We investigate the interplay between the Josephson energy, $E_\text{J}$, and the characteristic impedance, $Z_\text{c}$, in a unimon operated at its sweet spot of $\varphi = \pi$. We aim to understand how changes in these parameters influence the behavior of the system.

Starting with the auxiliary-mode Hamiltonian given by Eq.~\eqref{eq:H_aux}, we define new rescaled flux operators as $\hat{\tilde \Psi}= \sqrt{C_\text{eff}} \hat \Psi$ and $\hat{ \tilde \chi}_m = \sqrt{C_\text{eff}} \hat \chi_m $. The corresponding charge operators are defined as $\hat{\tilde Q}=  \hat Q / \sqrt{C_\text{eff}}$ and $\hat{ \tilde \Xi}_m = \hat \Xi_m / \sqrt{C_\text{eff}}$. This modification presents us with the Hamiltonian
\begin{align} \label{eq:H_aux_rescaled}
    \hat{H}_{\text{aux}} = &\frac{\hat{\tilde Q}^2}{2} + \frac{\hat{\tilde \Psi}^2}{2L_{\text{eff}} C_\text{eff}} + E_{\text{J}} \cos \biggl( \frac{2 \pi}{\Phi_0} \frac{\hat{\tilde \Psi}}{\sqrt{C_\text{eff}}} \biggr) \notag  \\
    & +\sum_{m = 1}^M \biggl[ \frac{\hat{\tilde \Xi}_m^2}{2} + \frac{1}{2} \Omega_m^2 \hat{\tilde \chi}_m^2 + \frac{\xi_m}{C_\text{eff}} \hat{\tilde \chi}_m \hat{\tilde \Psi} \biggr], 
\end{align}
where we have used effective inductance defined in Eq.~\eqref{eq:L_eff}.
We postulate that, given the high plasma frequency $\omega_\text{p} = 1/\sqrt{L_\text{J}C_\text{J}}$ compared to the frequencies of the primary normal modes, the Josephson capacitance $C_\text{J}$ in the effective capacitance can be reasonably neglected. This assumption is reinforced by our numerical findings.

The characteristic impedance $Z_\text{c}$ manifests in the relevant quantities as
\begin{align}
C_\text{eff} \propto Z_\text{c}^{-1}, \qquad
L_\text{eff} \propto Z_\text{c}, \qquad
\xi_m \propto Z_\text{c}^{-1}.
\end{align}
Applying these relations to Eq.~\eqref{eq:H_aux_rescaled}, it follows that the cosine term is the only one depending on 
$Z_\text{c}$. An expansion using a Taylor series provides
\begin{align}
E_{\text{J}} \cos \left( \frac{\hat{\tilde \varphi}}{\sqrt{C_\text{eff}}} \right) = &E_{\text{J}} - \beta_2 E_{\text{J}} Z_\text{c} \hat\varphi^2 +\beta_4 E_{\text{J}} Z_\text{c}^2 \hat\varphi^4 +... \notag
\end{align}
where $\beta_2$ and $\beta_4$ are constants, and $\hat{\tilde \varphi} = 2 \pi \hat{\tilde \Psi} / \Phi_0$.

This expansion underscores an essential observation, that $E_\text{J}$ and $Z_\text{c}$ influence the system in roughly identical ways up to the second order. However, the distinction emerges in higher-order terms. Notably, the sensitivity of anharmonicity to changes in $Z_\text{c}$ increases with increasing $Z_\text{c}$. Yet, one must also account for the renormalization effects from other modes. These effects, which become more pronounced at greater $Z_\text{c}$, act to temper the increase in anharmonicity.

In summary, our investigation reveals that, within the bounds of our assumptions, an equal relative change in either $E_\text{J}$ or $Z_\text{c}$ produces an identical outcome on the system up to linear order. The differences primarily arise in high-order behavior and in the distinct effects of renormalization at an elevated $Z_\text{c}$.

\section{Analytical calculations in the harmonic-oscillator basis} \label{app:Analytical calculations}

We begin with examining the cosine term in the normal-mode representation, which is the source of nonlinearity in our system: $E_\text{J}\cos \bigl(\sum_{m = 1}^{M+1} \hat{\varphi}_m  \bigr)$. Following the steps taken in Ref.~\cite{leib_networks_2012}, we neglect the transverse-type interactions, rapidly rotating terms by the rotating-wave approximation, and Kerr-type interactions operating on more than three modes. This allows us to express the cosine term as
\begin{widetext}
\begin{align}
E_\text{J}\cos \Biggl(\sum_{m = 1}^{M+1} \hat{\varphi}_m  \Biggr) &\approx E_\text{J}\prod_{m = 1}^{M+1} \ee^{-\lambda_m^2/2} \bigg( 1 - \lambda_m^2 \hat{a}^{\dagger}_m\hat{a}_m + \frac{\lambda_m^4}{4} \hat{a}^{\dagger}_m\hat{a}^{\dagger}_m \hat{a}_m\hat{a}_m + ... \bigg) \\ 
&\approx - E_\text{J}^* \sum_{m = 1}^{M+1} \Biggl( \lambda_m^2 \hat{a}^{\dagger}_m\hat{a}_m - \frac{\lambda_m^4}{4} \hat{a}^{\dagger}_m\hat{a}^{\dagger}_m\hat{a}_m\hat{a}_m + ... \Biggr) \Biggl[ 1 - \sum_{\substack{n = 1 \\ n \neq m}}^{M+1} \lambda_n^2 \hat{a}^{\dagger}_n\hat{a}_n  + \frac{1}{2} \sum_{\substack{k = 1 \\ k \neq m}}^{M+1}  \lambda_k^2 \hat{a}^{\dagger}_k\hat{a}_k  \sum_{\substack{l = 1 \\ l \notin \{ m,k \}}}^{M+1}  \lambda_l^2 \hat{a}^{\dagger}_l\hat{a}_l  \Biggr],  \notag
\end{align}
\end{widetext}
where we have used the harmonic-oscillator basis $\hat \varphi_m~=~\lambda_m \bigl( \hat a_m^\dagger + \hat a_m \bigr)$ and $E_\text{J}^* = E_\text{J}\prod_{m = 1}^{M+1} \ee^{-\lambda_m^2/2}$. From this result, we identify the self- and cross-Kerr interactions including high-order corrections for any three-mode combination as given in Eqs.~\eqref{eq:alpha_analytical} and~\eqref{eq:K_analytical} of the main text.

\bibliography{bibliography}% Produces the bibliography via BibTeX.

\end{document}